%% file: expander_walks_paper_draft.tex
\documentclass[11pt]{article}
\usepackage[left=1in, right=1in, top=1in, bottom=1in]{geometry}
\usepackage[utf8]{inputenc}
\usepackage{amsmath, mathtools, amssymb, xcolor}
\usepackage[colorlinks]{hyperref}
\usepackage[colorinlistoftodos]{todonotes}
\usepackage{graphicx}
\usepackage{subcaption}
\usepackage{amsthm}
\usepackage{outlines}
\usepackage{authblk}

\usepackage[nottoc,numbib]{tocbibind}

\PassOptionsToPackage{unicode}{hyperref}
\PassOptionsToPackage{naturalnames}{hyperref}

\usepackage{amsthm}

\newtheorem{theorem}{Theorem}[section]
\newtheorem{thrm}[theorem]{Theorem}
\newtheorem{cor}[theorem]{Corollary}
\newtheorem{lemma}[theorem]{Lemma}
\newtheorem{prop}[theorem]{Proposition}
\newtheorem{defn}[theorem]{Definition}

\newcommand\reallywidehat[1]{\arraycolsep=0pt\relax%
\begin{array}{c}
\stretchto{
  \scaleto{
    \scalerel*[\widthof{\ensuremath{#1}}]{\kern-.5pt\bigwedge\kern-.5pt}
    {\rule[-\textheight/2]{1ex}{\textheight}} 
  }{\textheight} %
}{0.5ex}\\           
#1\\                 
\rule{-1ex}{0ex}
\end{array}
}

\newcommand{\norm}{\|}
\newcommand{\arrbegin}{\begin{eqnarray}}
\newcommand{\arrend}{\end{eqnarray}}

\newcommand{\absl}{\left|}
\newcommand{\absr}{\right|}

\newcommand{\RR}{\mathbb{R}}
\newcommand{\CC}{\mathbb{C}}
\newcommand{\ZZ}{\mathbb{Z}}
\newcommand{\NN}{\mathbb{N}}
\newcommand{\EE}{\mathop{\mathbb{E}}}
\newcommand{\PP}{\mathop{\mathbb{P}}}
\newcommand{\FF}{\mathbb{F}}

\newcommand{\rar}{\rightarrow}

\newcommand{\suml}{\sum\limits}

\newcommand{\defeq}{\vcentcolon=}

\newcommand{\eps}{\epsilon}

\newcommand{\la}{\langle}
\newcommand{\ra}{\rangle}
\newcommand{\pll}{\parallel}
\newcommand{\bb}{\{0, 1\}}

\DeclarePairedDelimiter\floor{\lfloor}{\rfloor}
\DeclarePairedDelimiter\abs{\lvert}{\rvert}

\usepackage{amsfonts}
\usepackage{amsmath}
\usepackage{textcomp}
\usepackage{wasysym}
\begin{document}

\title{Near-Optimal Cayley Expanders for Abelian Groups}

\newcommand\CoAuthorMark{\footnotemark[\arabic{footnote}]} 

\author[1]{Akhil Jalan \footnote{This material is based upon work supported by the National Science Foundation under grants number 1218547 and 1678712.}}
\author[1]{Dana Moshkovitz \protect\CoAuthorMark}
\affil[1]{Department of Computer Science, UT Austin}
\affil[ ]{\textit {\{akhil,danama\}@cs.utexas.edu}}
\date{}                     
\setcounter{Maxaffil}{0}
\renewcommand\Affilfont{\itshape\small}

\maketitle

\input{abstract}

\newpage
{
  \hypersetup{linkcolor=black}
  \tableofcontents
}

\newpage
\input{intro}

\newpage
\input{preliminaries}

\newpage
\input{body}

\newpage
\input{applications}

\newpage
\bibliographystyle{alpha}
\bibliography{expander_walks}

\end{document}

%% file: abstract.tex

\begin{abstract}

We give an efficient deterministic algorithm that outputs an expanding generating set for any finite abelian group. The size of the generating set is close to the randomized construction of Alon and Roichman (1994), improving upon various deterministic constructions in both the dependence on the dimension and the spectral gap. By obtaining optimal dependence on the dimension we resolve a conjecture of Azar, Motwani, and Naor (1998) in the affirmative. Our technique is an extension of the bias amplification technique of Ta-Shma (2017), who used random walks on expanders to obtain expanding generating sets over the additive group of n-bit strings. As a consequence, we obtain (i) randomness-efficient constructions of almost k-wise independent variables, (ii) a faster deterministic algorithm for the Remote Point Problem, (iii) randomness-efficient low-degree tests, and (iv) randomness-efficient verification of matrix multiplication.

\end{abstract}

%% file: intro.tex
\newcommand{\poly}{\text{poly}}

\section{Introduction}

\subsection{Main Result}

A graph is an expander if there exists $\alpha > 0$ such that the spectral gap of its adjacency matrix (namely, the difference between its top eigenvalue and its second eigenvalue) is at least $\alpha$. Such graphs are very well-connected in the sense that they lack sparse cuts. Expanders that are additionally sparse are immensely important in computer science and mathematics (see, e.g. the survey \cite{hoory2006expander}). 

Cayley graphs are an important class of graphs built from groups. Given a group $G$ and a generating set $S \subset G$, the graph $Cay(G, S)$ has vertex set $G$ and edges $(g, g \cdot s)$ for all $g \in G$, $ s \in S$. In addition to describing various well-known graphs such as the hypercube and the torus, Cayley graphs of (non-abelian) groups gave the first explicit constructions of near-optimal expander graphs \cite{lps}. 
Moreover, their 
algebraic structure makes Cayley graphs easier to analyze. In particular, the eigenvectors and eigenvalues of a Cayley graph are well-understood through the Fourier transform on the group. 
 
When is a Cayley graph an expander? Alon and Roichman showed that given a group $G$, integer $n \geq 1$, and $\eps > 0$, taking a uniformly random subset $S \subset G^n$ of size $O(\frac{n \log(\abs{G})}{\eps^2})$ is an expander with spectral gap $1 - \eps$, with high probability \cite{alon-roichman}. They also proved a nearly matching lower bound of $\abs{S} = \Omega((\frac{n \log(\abs{G})}{\eps^2})^{1 - o(1)})$ when $G$ is abelian. When $G = \FF_2$ the lower bound is $\Omega(\frac{n}{\eps^2 \log(1/\eps)})$ \cite{AGHP} \footnote{It is possible that this lower bound is tight. A candidate construction based on algebraic-geometric codes could achieve this lower bound ~\cite{ben-aroya-ta-shma}.}.

An explicit construction with parameters matching the Alon-Roichman bound has remained elusive, despite being widely studied in the pseudorandomness literature \cite{katz-1989, naor-naor, AGHP, rsw, aik-thin-set, alon-mansour, EGL, amn, chen-moore-russell, arvind-rpp, ben-aroya-ta-shma, arvind-permutation-groups}. 

The best known results achieve $O((\log(\abs{G}) + \frac{n^2}{\eps^2})^{5})$ for arbitrary abelian $G$ \cite{arvind-rpp}, $O(\frac{n^2}{\eps^2})$ for abelian $G$ where $\abs{G} \leq \log(\frac{n^2}{\eps^2})^{O(1)}$, and $O(\frac{n \log(\abs{G})^{O(1)}}{\eps^{11}})$ for general $G$ \cite{chen-moore-russell}. For solvable subgroups of permutation groups one can improve this to $O(\frac{n^2}{\eps^{8}})$ ~\cite{arvind-permutation-groups}. 

In this paper we give an explicit construction of expanding generating sets for abelian groups whose size is near the Alon-Roichman bound. 

\begin{thrm}\label{main-thrm}
There is a deterministic, polynomial-time algorithm which, given a generating set of an abelian group $G$, integer $n \geq 1$, and $\eps > 0$, outputs a generating set $S \subset G^n$ of size $O(\frac{n \log(\abs{G})^{O(1)}}{\eps^{2 + o(1)}})$ such that $Cay(G^n, S)$ has spectral gap $1-\eps$. 
\end{thrm}

Expanding Cayley graphs are equivalent to pseudorandom objects called $\eps$-biased sets. These were originally defined over $\FF_2^n$ by Naor and Naor \cite{naor-naor}. A set $S \subseteq \FF_2^n$ is said to be $\eps$-biased if for every non-empty $T \subseteq [n]$, we have $\EE\limits_{x \in S} [\bigoplus\limits_{i \in T} x_i] = 1/2 \pm \eps$. 

Naor and Naor initiated a long line of work culminating in a recent breakthrough result by Ta-Shma, that achieves $\abs{S} = O(\frac{n}{\eps^{2 + o(1)}})$ \cite{ta2017explicit}. This construction approaches the Alon-Roichman bound as $\eps \to 0$.  

Ta-Shma's construction follows previous work in using a 2-step ``bias amplification'' approach. First, identify an explicit set $S_0 \subset \FF_2^n$ with constant bias, usually through algebraic methods. Second, amplify the bias of $S_0$ to any $\eps > 0$ by performing a random walk on an expander graph. While this general method was already known, it could only achieve $\abs{S} = O(\frac{n}{\eps^{4 + o(1)}})$. To break this barrier, Ta-Shma identified a graph structure obtained from a ``wide replacement product", which was more effective for the bias amplification step and resulted in $\abs{S} = O(\frac{n}{\eps^{2 + o(1)}})$. 

\subsection{Wide Replacement Walks are Optimal Character Samplers}

Random walks on expander graphs are useful for a variety of algorithmic purposes. A classical fact is that expander walks are good approximate samplers, in the sense that a sufficiently long random walk on an expander will visit sets of density $\delta$ for approximately a $\delta$ fraction of the steps. This is called the ``expander Chernoff bound'' and one can characterize this as the property that expander walks fool a suitable test function.  

Ta-Shma observed that expander walks fool the much more sensitive class of parity functions on $\bb^n$ as well. Parity functions are sensitive to input perturbations - flipping a single bit in the input can change the output. The classical expander Chernoff bound is not fine-grained enough to prove that $t$-step expander walks fool parity functions. The fact that they nevertheless do fool parity functions is therefore surprising, and Ta-Shma referred to this fact as ``expanders are good parity samplers'' \cite{ta2017explicit}. 

Since parity functions are just the characters of $\FF_2^n$, we can ask: do expander walks also fool the characters of more general classes of groups? We show that this is indeed true, and therefore ``expander walks are good character samplers.'' Moreover, just as in the $\FF_2$ case, a random walk on a wide replacement product of expander graphs is an optimal type of character sampler.

Therefore, wide replacement walks obtain precisely the same bias amplification parameters for small-bias sets over abelian groups as they do in the $G = \FF_2$ case. 

\textbf{Character sampling explained}: Let us precisely explain what we mean by ``character sampling.''  A character of an abelian group is a homomorphism $\chi: G \rar \CC^*$, where $\CC^*$ is the multiplicative group of complex numbers. The eigenvalues of an abelian Cayley graph $Cay(G, S)$ are given by $\abs{\EE_{x \sim S}\chi(x)}$ for all characters $\chi$. Note that the constant function that maps all values to $1$ is a character, and the eigenvalue associated with it is the top eigenvalue. Therefore, we are interested in generating sets $S$ such that $\abs{\EE_{x \sim S}\chi(x)} \leq \eps$ for all non-constant $\chi$. 

For simplicity, consider the case $G = \ZZ_d$ for some $d \geq 2$. Let $\omega_d \defeq exp(\frac{2\pi i}{d})$ denote the primitive $d^{th}$ root of unity. In this case the characters are just the maps $x \mapsto \omega_d^{x \cdot j}$ for $j = 0, 1, ..., d-1$. 

Now, suppose we have some $\eps_0$-biased set $G_0 \subset G$, where $\eps_0 < 1$ is a constant. First, observe that taking $t$ \textit{independent} samples from $G_0$ and outputting their sum obtains a distribution with bias $(\eps_0)^t$. However, since independent sampling also results in a distribution of size $\abs{G_0}^t$, there is no improvement in size as a function of bias. 

The idea of the random walk approach is to derandomize independent sampling by taking \textit{correlated} samples. Specifically, identify $G_0$ with the vertices of some degree-regular expander graph $\Gamma$. We need to show that taking a random walk of length $t$ on $\Gamma$ and then summing the elements in the path gives a distribution with less bias than $G_0$. 

A $t$-step walk on $\Gamma$ gives a sequence of group elements $(x_0, ..., x_t) \in G_0^{t+1}$. We are interested in the bias of the random group element $\sum_{i} x_i$. In general, we cannot hope that $(\sum_i x_i)$ is close to the uniform distribution in \textit{statistical distance}. However, for every non-constant character $\chi$, it turns out that the quantity $\abs{\EE[\chi(\sum_{i} x_i)]}$ is at most $\eps$, where the expectation is over paths $(x_0, ..., x_t)$ in the graph. Notice that $\EE_{x \in G} [\chi(x)] = 0$, so the random element $(\sum_i x_i)$ is close to uniform in the weaker sense of fooling characters. Therefore, the expander walk is a good ``character sampler.''


As in the $\FF_2$ case, character functions are sensitive to input perturbations. Therefore, the character sampling property of expanders is a much finer-grained property than the expander Chernoff bound. 


\textbf{Why expanders are character samplers}: We express the bias of the random walk distribution algebraically in terms of matrix norms corresponding to the random walk. 

Abusing notation, let $\Gamma$ denote the random walk matrix of the graph $\Gamma$. Let the character $\chi^*: \ZZ_d \rar \CC$ be the worst-case character for the random-walk distribution. Partition $G_0$ into $S_0, ..., S_{d-1}$ depending on their values with respect to $\chi^*$, so that $x \in S_k \iff \chi^*(x) = \omega_d^{k}$. 

We need to track how often the walk enters $S_0, S_1, ..., S_{d-1} \subset V(\Gamma)$. Identify each $S_i$ with an $\abs{S_i}$-dimensional subspace of $\CC^{V(\Gamma)}$. For $i \in \ZZ_d$ let $\Pi_i: \CC^{V(\Gamma)} \rar \CC^{V(\Gamma)}$ be the projection onto this subspace. Finally, let $\Pi = \sum_{y \in \ZZ_d} \omega_d^{y} \Pi_y$ be the weighted projection matrix. 

Given some initial distribution $\vec{u}$ on the vertices, the vector $\Gamma^{t} \vec{u}$ tracks the distribution after taking a $t$-step walk on the graph. The matrix $\Pi$ tracks how often the walk enters the sets $S_0, ..., S_{d-1}$, and so the bias of the random walk distribution can be bounded by the norm of $(\Pi \Gamma)^t$. 

Let $V^\pll$ denote the subspace spanned by the all-ones vector $\vec{1}$, and $V^\perp = (V^\pll)^{\bot}$. For a vector $v \in V^\pll \oplus V^\perp$, let $v^\pll$ and $v^\perp$ denote the projections onto $V^\pll, V^\perp$ respectively. 

While $\norm \Pi \Gamma \norm = 1$ since $\norm \Pi \Gamma \vec{1} \norm = \norm \Pi \vec{1} \norm = 1$, it turns out that $\norm (\Pi \Gamma)^2 \norm \leq bias(G_0) + 2 \lambda(\Gamma)$, where $\lambda(\Gamma)$ is the second eigenvalue of $\Gamma$ in absolute value. 

To see this, notice that if $\vec{v} \in V^\perp$ is a unit vector, then $\norm \Pi \Gamma \Pi \Gamma \vec{v} \norm \leq \norm \Pi \Gamma \Pi \norm \lambda(\Gamma) \norm \vec{v} \norm \leq \lambda(\Gamma)$. Therefore, the ``bad'' case is when $\vec{v} \in V^\pll$. Let $u = \frac{1}{\sqrt{\abs{V(\Gamma)}}} \vec{1}$. Using the fact that $\norm \Pi \norm = 1$, 

\begin{align}
\norm \Pi \Gamma \Pi \Gamma u \norm = \norm \Pi \Gamma \Pi u \norm \\
\leq \norm \Pi \Gamma (\Pi u)^\pll \norm + \norm \Pi \Gamma (\Pi u)^\perp \norm \\
\leq \norm \Pi (\Pi u)^\pll \norm + \lambda(\Gamma) \norm \Pi(\Pi u)^\perp \norm \\
\leq \norm \Pi (\Pi u)^\pll \norm + \lambda(\Gamma) 
\end{align}

It remains to show that $\norm \Pi (\Pi u)^\pll \norm \leq bias(G_0)$. To see this, notice that $\Pi$ is a diagonal matrix and $u$ is just $\vec{1}$ scaled by a constant. Further, $\Pi$ is a block-diagonal matrix of the form 

\begin{align}
\Pi = \begin{bmatrix}
I_{\absl S_0 \absr} & & & \\ & \omega_d I_{\absl S_1 \absr} & & \\ & & \ddots & \\ & & & \omega_d^{{d}-1} I_{\absl S_{{d}-1} \absr}
\end{bmatrix}
\end{align}

Note that we have reordered the vertices of the graph in order of $S_0, S_1$ and so on.  

If the blocks are exactly the same size, then $\Pi u \in V^\perp$, because $\sum_{y \in \ZZ_d} \omega_d^y = 0$. In general the blocks have different dimensions, but they are the same size up to the bias of $G_0$.  Therefore $\norm (\Pi u)^\pll \norm \leq bias(G_0)$. 


It follows that a random walk on $\Gamma$ is a good character sampler. However, this approach can never amplify bias fast enough to achieve a generating set smaller than $O(\frac{\abs{G_0}}{\eps^{4 + o(1)}})$. The reason is because while we can bound $\norm (\Pi \Gamma)^2 \norm$, we cannot bound $\norm \Pi \Gamma \norm$ below $1$. Therefore, we effectively only gain from one in every two steps. 

\newcommand{\HH}{\dot{H}}
\newcommand{\GG}{\dot{G}}
\newcommand{\PPi}{\dot{\Pi}}
\newcommand{\Ga}{\dot{\Gamma}}

\textbf{Wide Replacement Walks are Optimal Character Samplers}: To circumvent the ``2-step barrier'' of expander walks outlined above, Ta-Shma used the \textit{wide replacement walk} on a product of two expander graphs. The idea of the wide replacement walk is to take the product of a $D_1$-regular graph $\Gamma$ as before with an ``inner graph'' $H$ on $D_1^s$ vertices, for some $s \geq 2$. The product graph replaces every vertex of $\Gamma$ with a copy of $H$ (called a ``cloud'') and then connects clouds to other clouds according to the edge structure of $\Gamma$.  

Analyzing the bias of the walk involves bounding the matrix norm of $\PPi \Ga \HH$, where $\Ga$ and $\HH$ are random walk matrices on the product corresponding to $\Gamma, H$. 

\newcommand{\rrp}{\raisebox{.5pt}{\textcircled{\raisebox{-.9pt} {r}}}}

Let $V^\pll$ denote the subspace of vectors which are constant on the $H$-component of the product, and let $V^\perp = (V^\pll)^\perp$. 

Similar to the above case, one can show that $\PPi \Ga \HH$ shrinks the norm of any $v \in V^\perp$ by a factor of $\lambda(H)$. The difficult case is when $v \in V^\pll$. Here we arrive at the core idea of the replacement product: if the inner graph $H$ is \textit{pseudorandom} with respect to $\Gamma$, then when the walk is in $V^\pll$, the next $s$ steps approximate the ordinary random walk on $\Gamma$. 

This is enough to circumvent the ``2-step barrier'' since in even the ``bad case'' where the walk is stuck in $V^\pll$, we can shrink the bias as though it were taking an ordinary walk on $\Gamma$. As we showed above, this shrinks the bias from some $\eps_0$ to $(\eps_0 + 2\lambda(\Gamma))^{\floor{s/2}}$ every $s$ steps. If we select $\Gamma, H$ such that $\eps_0 + 2\lambda(\Gamma) \leq \lambda(H)^2$, then we conclude that we shrink the bias by a factor of $\lambda(H)^{s - O_s(1)}$ every $s$ steps. So we gain from $s - O(1)$ out of every $s$ steps.  

Going from the $\FF_2$-case to the case of general abelian groups simply requires a more careful analysis of characters. Morally speaking, the only difference in the analysis is that the projection matrix $\Pi$ which tracks how often the walk enters each $S_i$ is different. This does not change the overall argument much; in particular, we can use almost identical graphs $\Gamma, H$ as in \cite{ta2017explicit}. 

We conclude that a wide replacement walk allows us to amplify bias of a constant-biased subset $G_0 \subset G^n$ of size $O(n \log(\abs{G})^{O(1)})$ (e.g. the construction of \cite{arvind-permutation-groups}) to an $\eps$-biased set of size $O(\frac{n \log(\abs{G})^{O(1)}}{\eps^{2 + o(1)}})$, nearly matching the Alon-Roichman bound. 

\subsection{Applications} 

Explicit constructions of expander graphs are an essential component of algorithms, especially for derandomization. Here we are interested in the setting of constructing an expanding Cayley graph from a given abelian group $G$. Our construction achieves a near-optimal degree, which improves parameters in various applications. 

\textbf{Almost $k$-wise independence}: A distribution $D \sim G^n$ is $(\eps, k)$-wise independent if for every index set $I \subset [n]$ of size $k$, the restriction of $D$ to $I$ is $\eps$-close to uniform in statistical distance. Almost $k$-wise independent distributions are a fundamental object in and of themselves. They also have a variety of applications in derandomization, including load balancing \cite{christiani2014generating}, derandomization of Monte-Carlo simulations \cite{christiani2014generating}, derandomization of CSP approximation algorithms \cite{charikar-csp}, and pseudorandom generators \cite{chhl}. We note that certain applications (e.g. quantum $t$-designs \cite{ambainis-designs}) really require almost $k$-wise independent distributions over \textit{arbitrary} alphabet size rather than just the binary alphabet, which motivates our study of $\eps$-biased sets over arbitrary abelian groups. 

Vazirani's XOR Lemma asserts that an $\eps$-biased distribution $D$ is also $(\eps \sqrt{\abs{G}^k}, k)$-wise indepdent for all $k \leq n$. Therefore, by constructing an $\eps^\prime$-biased distribution where $\eps^\prime = \frac{\eps}{\sqrt{\abs{G}^k}}$, we also obtain explicit constructions of $(\eps, k)$-wise independent random variables on $G^n$. 

\begin{prop}[Almost $k$-wise independent sets over abelian groups]
Let $G$ be a finite abelian group given by some generating set. For any $\eps > 0$ and $n \geq k \geq 1$ there exists a deterministic, polynomial-time algorithm whose output is an $(\eps, k)$-wise independent distribution over $G^n$. The support size is $O(\frac{n \cdot \abs{G}^{k + o(1)}}{\eps^{2 + o(1)}})$. 
\end{prop} 

\textbf{Remote Point Problem}: A matrix $A \in \FF_2^{m\times n}$ is $(k, d)$-rigid iff for all rank-$k$ matrices ${R \in \FF_2^{m\times n}}$, the matrix $A - R$ has a row with at least $d$ nonzero entries. Valiant initiated the study of rigid matrices in circuit complexity, proving that an explicit construction of an $(\Omega(n), n^{\Omega(1)})$-rigid matrix for $m = O(n)$ would imply superlinear circuit lower bounds \cite{valiant1977}. After more than four decades of research, state of the art constructions have yet to meet this goal \cite{bhangale2020rigid}. 

The Remote Point Problem was introduced by Alon, Panigrahy, and Yekhamin as an intermediate problem in the overall program of rigid matrix constructions \cite{alon-2009-rpp}. Arvind and Srinivasan generalized the problem to any group \cite{arvind-rpp}. 

Let $G$ be a group, $n \geq 1$, and $H \leq G^n$ a subgroup given by some generating set. For a given $G, H$ and integer $r > 0$, the Remote Point Problem is to find a point $x \in G^n$ such that $x$ has Hamming distance greater than $r$ from all $h \in H$, or else reject. In the case of $G^n = \FF_2^n$, this is a relaxation of the matrix rigidity problem, since rather than finding $m$ vectors $x_1, ..., x_m \in \FF_2^n$ whose linear span is far from all low-dimensional subspaces, we are given a single subspace and must find just a single point far from it. 

To find a remote point, existing algorithms first construct a collection of subgroups $H_1, ..., H_m \leq G^m$ whose union covers all points of distance at most $r$ from $H$. In the $\FF_2$ case, \cite{alon-2009-rpp} find a point $x \not \in \bigcup\limits_{i} H_i$ by the method of pessimistic estimators. In the general case, \cite{arvind-rpp} instead prove that any generating set $S \subset G^n$ such that $Cay(G^n, S)$ has sufficiently good expansion must contain a point outside of $\bigcup\limits_{i} H_i$. They find this remote point by first constructing an expanding generating set $S$, and then exhaustively searching it. Their argument implicitly uses the fact that small-bias sets correspond to rigid matrices, albeit with weak parameters - this connection was developed further in \cite{alon-cohen-rigid-matrices}. 

The construction of \cite{arvind-rpp} for small-bias sets over abelian groups has size $O((\log(\abs{G}) + \frac{n^2}{\eps^2})^5)$ in general, and for $\log(\abs{G}) \leq \log(\frac{n^2}{\eps^2})^{O(1)}$ this is improved to $O(\frac{n^2}{\eps^2})$. Our algorithm improves the dependence on $n$ from $n^2$ to $n$.   



\newcommand{\lines}{\mathbb{L}}

\textbf{Randomness-Efficient Low-Degree Testing}: Let $\FF_q$ be the finite field on $q$ elements. Low-degree testing is a property testing problem in which, when given query access to a function $f: \FF_q^n \rar \FF_q$ and $d \geq 1$, one must decide whether $f$ is a degree $d$ polynomial or far (in Hamming distance) from all degree $d$ polynomials. These tests are a key ingredient in constructions of Locally Testable Codes (LTCs) and Probabilistically Checkable Proofs (PCPs) \cite{vadhan-pcp}.  

To test whether $f$ is a degree-$d$ polynomial, a natural test is to sample $x, y \sim \FF^n$ and check whether $f(x)$ agrees with the unique (degree-$d$, univariate) polynomial obtained by Lagrange interpolation along $d+1$ points on the line $\{x + ty: t \in \FF_q\}$.

Rubinfeld and Sudan introduced a low-degree test using this idea \cite{rubinfeld-sudan-1996}. It is given query access to the function $f$, along with a \textit{line oracle} function $g$. Let $\lines$ denote all lines $\{\vec{a} + t\vec{b}: t \in \FF_q\} \subset \FF_q^n$, where $\vec{a}, \vec{b} \in \FF^n$. Given a description of a line, the line oracle $g$ returns a univariate polynomial of degree $d$ defined on that line. Hence we write $g: \lines \rar \FF_q[t]$, where the image of $g$ is understood to only contain degree-$d$ polynomials. 

If $f$ is indeed a degree-$d$ polynomial, then one can set $g(\ell) = f\vert_{\ell}$ for all $\ell \in \lines$, and the following two-query test clearly accepts. 

(i) Select $x, y \in \FF^n$ independently, uniformly at random.

(ii) Let $\ell$ be the line determined by $\{x + ty: t \in \FF\}$. Accept iff $f(x)$ agrees with $g(\ell)(x)$. 

They also showed this test is sound: when $f$ is far from degree-$d$ polynomials, the test rejects with high probability. 

Ben-Sasson et al derandomized this test by replacing the second uniform sample $y$ with a sample from an $\eps$-biased set \cite{vadhan-pcp}. This modification improves the randomness efficiency of the tests, and therefore the length of the resulting LTC and PCP constructions. Moreover, they showed that the soundness guarantees of low-degree tests are almost unchanged due to the expansion properties of the Cayley graph on $\FF_q^n$.  

Our constructions of small-bias sets immediately imply improved randomness-efficiency of this low-degree test.  

\begin{prop}[Improved \cite{vadhan-pcp} Theorem 4.1]
Let $\FF_q$ be the finite field of $q$ elements, $n \geq 1$, $f: \FF_q^n \rar \FF_q$ a function, and $g: \lines \rar \FF_q[t]$ a line oracle. There exists a degree-$d$ test which has sample space size $O(q^n \cdot \frac{n\log(q)^{O(1)}}{\eps^{2 + o(1)}})$. For $d \leq q/3$ and sufficiently small $\delta > 0$, if the test accepts with probability $\geq 1- \delta$ then $f$ has Hamming distance at most $4\delta$ from a degree $d$ polynomial. 
\end{prop}

\textbf{Randomness-Efficient Verification of Matrix Multiplication}: Let $R$ denote some finite field $\FF_q$ or cyclic group $\ZZ_q$ for $q \geq 2$. Given  $A, B, C \in R^{n \times n}$, the matrix multiplication verification problem asks whether $AB = C$. 

Naively, one could multiply $A, B$ and then check whether $AB = C$ entry-wise in $O(n^{\omega})$ time, where $\omega \approx 2.373$ \cite{laser-2021}. A classical result of Freivalds suggests the following much simpler quadratic-time randomized algorithm: Sample $x \in R^n$ and check whether $ABx = Cx$ \cite{freivalds1977}. 

Observe that the entries of $ABx$ and $Cx$ are linear functions of $x$. Therefore, sampling $x$ from a small-bias set gives a randomness-efficient version of Freivalds' algorithm, at the cost of slightly higher error. Our construction therefore gives the following randomness efficient algorithm for verification of matrix multiplication. 

\begin{prop}
Let $R$ denote a finite field $\FF_q$ or cyclic group $\ZZ / q\ZZ$. Given matrices $A, B, C \in R^{n \times n}$ and $\eps$-biased set $S \subset R^n$, there exists randomized algorithm to decide whether $AB = C$ with one-sided error $(\frac{1}{q} + \eps)$. Its runtime is $O(n^2)$ and it uses $\log(\frac{n\log(q)^{O(1)}}{\eps^{2 + o(1)}})$ random bits.
\end{prop}

We note that if $R = \ZZ$, there exists a deterministic $O(n^2)$ time algorithm to verify matrix multiplication \cite{korec-wiedermann-2014}. However, this result relies on the fact that $\ZZ$ has characteristic zero. For the analysis to hold in the case of $\ZZ_q$, we would need a very strong bound on the entries of $A, B, C$ - namely, that $\max\limits_{i, j} \{\abs{A_{i, j}}, \abs{B_{i, j}}, \abs{C_{i, j}} \} \leq q^{\frac{1}{n-1}}$.

\subsection{Related Work}

\textbf{Explicit Constructions}: Explicit constructions of expanding generating sets for Cayley graphs have been mostly studied in the pseudorandomness literature in the context of small-bias sets for derandomization. Naor and Naor gave a combinatorial construction over $\FF_2^n$ of size $O(\frac{n}{\eps^3})$ \cite{naor-naor}. Alon, Goldreich, Hastad, and Peralta used algebraic arguments to give constructions over finite fields $\FF^n$ of size $O(\frac{n^2}{\eps^2})$, assuming the field size is bounded as $\log(\abs{\FF}) < \frac{n}{\log(n) + \log(1/\eps)}$ \cite{AGHP}. 

Resarchers in various communities have obtained constructions that achieve size $O(\poly(\frac{n \log(\abs{G})}{\eps}))$, but suboptimal exponents. In number theory and additive combinatorics researchers studying the case of $n = 1$ gave constructions over $\ZZ_{d}$ of size $O((\frac{\log(d)}{\eps})^{O(1)})$ \cite{rsw}, $O(\frac{\log(d)^{O(1)}}{\eps^2})$ \cite{katz-1989}, and $O(\frac{d}{\eps^{{O(\log^*(d))}}})$ \cite{aik-thin-set}. 

Other constructions equivalent to small-bias sets include $O(\frac{(n-1)^2}{\eps^2})$-sized $\eps$-discrepancy sets over finite fields of prime order $p$ when $n \leq p$ \cite{alon-mansour}, and $\eps$-balanced codes over finite fields, corresponding to small-bias sets over $\FF_q^n$ of size $O(n \cdot q)$ with constant bias \cite{justesen1972}. 

Ta-Shma's tour de force gave the first explicit construction of expanding generating sets of size $O(\frac{n\log(\abs{G})}{\eps^{2+o(1)}})$, nearly attaining the Alon-Roichman bound, but only for the special case of $G = \FF_2$ \cite{ta2017explicit}. Our work is an extension of Ta-Shma's bias amplification technique to the more general setting of arbitrary abelian groups. 

Azar, Motwani, and Naor generalized the study of small-bias sets to finite abelian groups \cite{amn}. Over $\ZZ_d^n$ they used character sum estimates to give a construction of size $O((d + \frac{n^2}{\eps^2})^C)$, where $C \leq 5$ is Linnik's constant \cite{linnik-2011-ref}. Assuming the Extended Riemann Hypothesis, ${C \leq 2 + o(1)}$ \cite{linnik-constant-riemman}. When $\log(d) \leq \log(\frac{n^2}{\eps^2})^{O(C)}$ they improve the size to $O((1 + o(1)) \frac{n^2}{\eps^2})$. 

Arvind and Srinivasan proved that one can project small-bias sets over $\ZZ_d^n$ to any abelian group $G^n$ when $d$ is the largest invariant factor of $G$. Therefore, using the construction from \cite{amn} they obtain small-bias sets over $G^n$ with the same bias and size as \cite{amn}, with $d = O(\log(\abs{G}))$ \cite{arvind-rpp}. 

The most general setting is to consider Cayley graphs over non-abelian groups. Wigderson and Xiao derandomized the Alon-Roichman construction using the method of pessimistic estimators \cite{wigderson-xiao-2008}. Arvind, Mukhopadhyay, and Nimbhorkhar later gave a derandomization for both directed and undirected Cayley graphs using Erdos-Renyi sequences \cite{arvind-2012-erdos-renyi}. 
However, both algorithms require the entire group table of $G^n$ as input, rather than just a generating set. Since generating sets are of size $O(n \log(\abs{G}))$, these algorithms are exponentially slower, running in time $O(\poly(\abs{G}^n))$ rather than $O(\poly(n \log(\abs{G}))$. Nevertheless, they have applications to settings such as homomorphism testing \cite{shpilka-wigderson-2006}, which Wigderson and Xiao derandomized using their construction of expanding generating sets \cite{wigderson-xiao-2008}. 

Chen, Moore, and Russell obtained generating sets of size $O(\frac{n}{\eps^{11}})$ over arbitrary groups $G^n$ where $\abs{G}$ is a constant \cite{chen-moore-russell} . Like Ta-Shma, their technique is to use bias amplification via expander graphs; specifically, they amplify bias via an iterated application of a 1-step random walk on an expander graph. Rozenmann and Wigderson had already noted that this technique amplifies bias for $G = \FF_2$ \cite{bogdanov-notes-rozenmann-wigderson}. Chen, Moore, and Russell generalized this analysis to all groups, using techniques from harmonic analysis and random matrix theory \cite{chen-moore-russell}. 

Existing work seems far from obtanining constructions for non-abelian groups near the Alon-Roichman bound. Known work tends to concentrate on special classes of non-abelian groups with some useful algebraic structure. Chen, Moore, and Russell constructed generating sets of size $O(\frac{(n\log(\abs{G}))^{1 + o(1)}}{\eps^{O(1)}})$ for smoothly solvable groups with constant-exponent abelian quotients \cite{chen-moore-russell}. Their analysis exploits the structure of solvable groups via Clifford theory. It also hinges on the assumption that the quotients in the derived series have constant exponent. 

Arvind et al later gave a construction of size $\Tilde{O}(\frac{\log(\abs{G})^{2 - o(1)}}{\eps^8})$ for solvable subgroups $G$ of permutation groups \cite{arvind-permutation-groups}. Their construction recursively generates expanding generating sets for quotients in the derived series of the group, and uses the thin sets construction of \cite{aik-thin-set} as a base set. Unlike \cite{chen-moore-russell} they do not require successive quotients of the derived series to be small; however, their argument does rely on an $O(\log(n))$ upper bound on the length of the derived series for any solvable $G \leq S_n$, which is not true for solvable groups in general. 

\textbf{Lower Bounds}: Alon and Roichman gave a randomized upper bound of $O(\frac{n\log(\abs{G})}{\eps^2})$ on the size of a generating set for any finite $G^n$ with spectral gap $(1-\eps)$ \cite{alon-roichman}. In the same paper, they gave a nearly matching lower bound when $G$ is abelian, of $\Omega((\frac{n\log(\abs{G})}{\eps^2})^{1 - o(1)})$. This is a sharper version of the folklore result that an abelian group $G^n$ requires $O(n \log(\abs{G}))$ generators for its Cayley graph to be connected.

For non-abelian groups, the existence of sparse expanders means the best lower bound in general is the Alon-Boppana bound. This removes the dependence on $\abs{G}$ and $n$, only requiring a generating set of size $\Omega(\frac{1}{\eps^2})$ \cite{alon-boppana} to achieve spectral gap of $1 - \eps$. Indeed, explicit constructions of Ramanujan graphs can be built from Cayley graphs of non-abelian groups \cite{lps}, and therefore attain this bound. 


\textbf{Expander Walks}: Random walks on expander graphs are an essential tool in computer science. Rather than surveying the vast literature, we refer the reader to the surveys \cite{hoory2006expander, vadhan-book}. Two remarks are in order. 

First, our use of wide replacement walks is essentially a way of building expander graphs from other expander graphs. This is thematic of several previous works, such as the zig-zag product \cite{rvw-zigzag-2000}. Note that the zig-zag product is just a modification of the replacement product; indeed, the (wide) replacement product itself can be used to give explicit, combinatorial constructions of Ramanujan graphs \cite{bats-ramanujan-graphs}. Ta-Shma used wide replacement walks to amplify spectral gaps of Cayley graphs on $\FF_2^n$ \cite{ta2017explicit}; this construction relied on previous constructions of expander graphs, although the expander graphs were not required to be Cayley graphs themselves. 

Second, the fact that ``expanders are good character samplers'' is surprising given that characters are sensitive to input perturbations. A recent work of Cohen, Peri, and Ta-Shma uses Fourier-analytic techniques to classify a large class of Boolean functions which can be fooled by expander walks, including all symmetric Boolean functions \cite{cohen2020expander}. 

\subsection{Open Problems}

\textbf{Expanding generating sets of optimal size}: The Alon-Roichman theorem proves that every group $G^n$ has an expanding generating set $S \subset G^n$ of size $\abs{S} = O(\frac{\log(\abs{G})}{\eps^2})$ \cite{alon-roichman}. This construction has not been fully derandomized for any group; even in the case of $G^n = \FF_2^n$, Ta-Shma's construction only asympotically approaches a size of $O(\frac{n}{\eps^{2}})$ as $\eps \to 0$. The actual size of the generating set is $O(\frac{n}{\eps^{2+o(1)}})$, and this $o(1)$ term is seemingly unavoidable when using expander walks \cite{ta2017explicit}. 

Similarly, our algorithm gives an expanding generating $S \subset G^n$ of size $O(\frac{n \log(\abs{G})^{O(1)}}{\eps^{2+o(1)}})$, for finite abelian $G$. The additional $\poly\log(\abs{G})$ factor comes from the bounds on constant-bias subsets of abelian groups; any construction of a constant-bias set $S \subset G^n$ of size $O(n \log(\abs{G}))$ would immediately give expanding generating sets of size $O(\frac{n \log(\abs{G})}{\eps^{2+o(1)}})$. To our knowledge, not even a candidate construction exists which would give constant-bias subsets of size $O(n\log(\abs{G}))$ for abelian groups; this is an interesting and potentially easier open problem, since it requires none of the expander walks machinery that we need to get arbitrarily small $\eps$. 

There is a candidate construction that could beat the Alon-Roichman bound for $G = \FF_2$, based on algebraic-geometric codes \cite{ben-aroya-ta-shma}. The code construction would give an $\eps$-biased set $S \subset \FF_2^n$ of size $\abs{S} = O(\frac{n}{\eps^2 \log(1/\eps)})$, assuming a conjecture in algebraic geometry. The authors themselves note that they have ``no idea'' whether this conjecture is valid \cite{ben-aroya-ta-shma}. 

\textbf{Expanding generating sets of non-abelian groups}: While wide replacement walks amplify bias quite naturally for abelian groups, it is unclear whether they can do so for general groups. Dealing with matrix-valued irreducible representations, rather than scalar-valued characters, makes the analysis of bias amplification considerably more involved; hence even the analysis of the 1-step walk is nontrivial \cite{chen-moore-russell}. It would be very interesting to see whether one can place algebraic conditions on a group that are weaker than commutativity, but still ensure that the wide replacement walk amplifies bias. 

Existing works on expanding generating sets for non-abelian groups have studied solvable groups, which generalize abelian groups \cite{chen-moore-russell, arvind-permutation-groups}. However, if we restrict the algorithm to input instances which are all non-abelian groups, then existence results suggest that one should be able to \textit{beat} the Alon-Roichman bound. 

For example, it is known that for every finite \textit{simple} non-abelian group $G^n$, there exists a generating set $S \subset G^n$ such that $Cay(G^n, S)$ has spectral gap $1-\eps$, and $\abs{S}$ is independent of $n$ \cite{breuillard-lubotzky-2018expansion}. Therefore, restricting input instances to simple groups seems too easy, while an algorithm for all groups seems too hard. Is there some natural natural class of non-abelian, non-simple groups for which algorithms can efficiently find expanding generating sets near (or even below) the Alon-Roichman bound? 

\textbf{Decoding over any finite field}: A recent work of Jeronimo et al gives a decoding algorithm for a modified version of Ta-Shma's codes \cite{jeronimo2020-decoding}. Since our work gives $\eps$-balanced codes over any finite field, it would be interesting to extend both the modification of the codes and the decoding algorithm of \cite{jeronimo2020-decoding} to this general setting. 

\textbf{Classifying the power of expander walks on groups}: So far we have discussed how random walks on expanders are good samplers in various ways, such as the expander Chernoff bound, parity sampling, and character sampling. Cohen, Peri, and Ta-Shma study the class of all Boolean functions that expander walks fool \cite{cohen2020expander}. It would be very interesting to extend their results to functions on groups, perhaps using similar tools from harmonic analysis and representation theory. For example, for which groups $G$ besides $\FF_2$ do expander walks fool all symmetric functions on $G^n$? 

%% file: preliminaries.tex

\section{Preliminaries}

\subsection{Cayley Graphs and Expanders}

In this paper we are concerned with the expansion of a particular kind of graph called a Cayley graph. We begin with some preliminaries on graphs and group theory. 

\begin{defn}[Spectral expander graph]
Let $G = ([n], E, w)$ be a weighted, $d$-regular undirected graph. By $d$-regular we mean that for all $u \in V$, $\sum_{v \in V} w(\{u, v\}) = d$. 

Let $A \in \CC^{n \times n}$ be the (weighted) adjacency operator of $G$, and let $M = \frac{1}{d} A$ be the normalized adjacency operator, also known as the random walk matrix. Let the eigenvalues of $M$ be denoted $\lambda_n \leq ... \leq \lambda_2 \leq \lambda_1 = 1$, counting multiplicity. Then $G$ is a one-sided spectral expander if $\lambda_2 < 1 - \Omega(1)$, and $G$ is a two-sided spectral expander if $$\max\{\absl \lambda_n \absr, \absl \lambda_2\absr \} < 1 - \Omega(1)$$ 

Let $\lambda(G) \defeq \max\{\absl \lambda_n \absr, \absl \lambda_2\absr \}$. The two-sided spectral gap of $G$ is $1 - \lambda(G)$.
\end{defn}

Throughout this paper, when speaking of expander graphs we will mean two-sided spectral expanders. We will commonly  use $\lambda(G)$ to denote the second eigenvalue in absolute value of a graph $G$. 

Next, we define Cayley graphs, which are a type of graph whose vertices correspond to elements of some group, and whose edges are defined by the group operation. 

\begin{defn}(Symmetric generating set)
Let $G$ be a group and $S \subset G$. We say that $S$ is symmetric if for all $s \in S$, $s^{-1} \in S$. Further, $S$ is a generating set if for all $g \in G$ there exist $s_1, ..., s_k \in S$ (possibly repeated) such that $$s_k \cdots s_1 = g$$

We write $\la S \ra = G$. 
\end{defn}

\begin{defn}(Cayley Graph)
Let $G$ be a group and $S \subset G$ be a symmetric generatring set, and $w: S \rar \RR_{\geq 0}$ a weight function. The Cayley graph $Cay(G, S, w)$ is the graph with vertex set $G$ and edge set $\{\{g, g \cdot s\}: g \in G, s \in S\}$. The weight of an edge $\{g, g \cdot s\}$ is $w(s)$. 
\end{defn}

We will require the total weight of $S$ to be normalized to $\abs{S}$ by convention. Notice that since $S$ is symmetric, we can consider the graph $Cay(G, S)$ to be an undirected and weighted $\abs{S}$-regular multigraph. 

The eigenvectors of abelian Cayley graphs are described by their characters. 

\begin{defn}
Let $\CC^*$ be the multiplicative group of nonzero complex numbers. For any finite abelian group $G$, the characters of $G$, denoted $\hat{G}$, are the set of all homomorphisms $\chi: G \rar \CC^*$. 
\end{defn}

\begin{prop}
Let $G$ be a finite abelian group and $S \subset G$ a symmetric generating set. Then the eigenvalues of $Cay(G, S)$ are given by 

$$\{\abs{\EE_{x \sim S}[\chi(x)]}: \chi \in \hat{G}\}$$
\end{prop}

We remark that one can generalize this definition to non-abelian groups by replacing characters with equivalence classes of irreducible unitary representations - see, e.g. \cite{chen-moore-russell}. For abelian groups the only such representations are character functions, so it suffices to discuss characters for our purposes. 

Notice that any group has a \textit{trivial character} $\chi: G \rar \CC^*$ such that $\chi(g) = 1$ for all $g$. The eigenvalue corresponding to the trivial character is always $1$. Therefore, for a Cayley graph to be an expander we need bounds on all of its nontrivial characters. This leads to the definition of an expanding generating set for an abelian Cayley graph, which is also known as a small-bias set. 

\begin{defn}[Small-bias distributions for abelian groups]
Let $G$ be a finite abelian group and $D \sim G$ a random variable. For any character $\chi$ of $G$, the bias of $D$ with respect to $\chi$ is 

$$Bias_{\chi}(D) \defeq \abs{\EE_{x \sim D}[\chi(x)]}$$

Let $\chi_0$ denote the trivial character. The \textit{bias} of $D$ is its maximum bias with respect to nontrivial characters. 

$$Bias(D) \defeq \max\limits_{\chi \neq \chi_0} Bias_{\chi}(D)$$

If $S \subset G$, then $bias(S)$ is the bias of the uniform distribution on $S$. If $S$ is a symmetric generating set, $\lambda(Cay(G, S)) = Bias(S)$. 
\end{defn}

Notice that if $S$ is non-negatively weighted, we can normalize weights to sum to $1$ and obtain a (not necessarily uniform) distribution on $S$. Then the bias of $S$ is just the bias of this distribution. 

In this language, the Alon-Roichman Theorem asserts that a random subset of $G$ of size $O(\frac{\log(\abs{G})}{\eps^2})$ is $\eps$-biased with high probability \cite{alon-roichman}. 

Finally, we will need a few more facts about characters of abelian groups. 

\begin{prop}(Characters of cyclic groups)
Let $\ZZ_d$ be the cyclic group on $d \geq 2$ elements. Let $\omega_d \defeq exp(\frac{2 \pi i}{d})$. The characters of $\ZZ_d$ are the maps $\chi_j(x) = \omega_d^{j \cdot x}$ for $j = 0, 1, ..., d-1$. 
\end{prop}

\begin{defn}(Direct sum of groups)
Let $A, B$ be abelian groups. The direct sum $A \oplus B$ is the abelian group whose elements belong to the Cartesian product $A \times B$. For $(a_1, b_1), (a_2, b_2) \in A \times B$, the group operation is

$$(a_1, b_1) + (a_2, b_2) = (a_1 + a_2, b_1 + b_2)$$
\end{defn}

Notice that the direct sum is associative. For abelian groups $A, B, C$, $(A \oplus B) \oplus C \cong A \oplus (B \oplus C)$. So we can write $A \oplus B \oplus C$ without ambiguity. 

\begin{prop}(Fundamental theorem of finite abelian groups)
Let $G$ be a finite abelian group. Then $G$ is isomorphic to a direct sum of cyclic groups. That is, there exist $d_1, ..., d_k \geq 2$ such that 

$$G \cong \ZZ_{d_1} \oplus \cdots \oplus \ZZ_{d_k}$$

Moreover, $d_i | d_j$ for all $i < j$. 

We refer to $\ZZ_{d_1} \oplus \cdots \oplus \ZZ_{d_k}$ as the invariant factor decomposition of $G$. The integers $d_1, ..., d_k$ are the invariant factors. 
\end{prop}

From the above propositions one can show that the characters of a finite abelian group are products of maps of the form $x \mapsto \omega_{d_i}^{j \cdot x}$. This structure is crucial to our overall argument. 

As a special case, consider $G = \FF_2^n$. The characters of $\FF_2^n$ are precisely the elements of the Fourier basis for the vector space of functions $\{f: \{-1, 1\}^n \rar \CC\}$. Fix any $T \subset [n]$. The character $\chi_T: \{-1, 1\}^n \rar \CC$ is given by the parity function on $T$. $$\chi_T(x) = \prod\limits_{i \in T} x_i$$

The trivial character corresponds to $T = \emptyset$. From this it is easy to see that an $\eps$-biased set $S \subset \FF_2^n$ is also balanced, in the sense that its expected parity on any non-empty substring is close to $1/2$. Equivalenty, $S \subset \FF_2^n$ is $\eps$-biased iff its indicator function $1_S: \{-1, 1\}^n \rar \{0, 1\}$ has bounded Fourier coefficients $\abs{\widehat{1_S}(T)} \leq \eps$ for all nonempty $T \subset [n]$. 

\subsection{Wide Replacement Walks}

\renewcommand{\rrp}{\raisebox{.5pt}{\textcircled{\raisebox{-.9pt} {r}}}}

Our algorithm performs a random walk on a wide replacement walk of expander graphs. In this section we define what it means to take a wide replacement walk. 

Let $G$ be a $D_1$-regular graph on $N_1$ vertices and $H$ be a $D_2$-regular graph on $D_1$ vertices. The \textit{replacement product} $G \rrp H$ is a $(D_2 + 1)$-regular graph on $N_1 \cdot D_1$ vertices. Each vertex of $G$ (the ``outer graph'') is replaced by a copy of $H$ (the ``inner graph''). We call these copies \textit{clouds}. 

The intra-cloud edges in each cloud of $G \rrp H$ are just the edges from $H$. However, $G \rrp H$ also has \textit{inter}-cloud edges which arise by identifying the $D_1$ vertices of $H$ with the $D_1$ incident edges of a vertex $v \in V(G)$. This identification requires that we number the edges of every vertex in $G$. We formalize this with the concept of a rotation map. 

\begin{defn}(Rotation map)
Let $G$ be a $D$-reguluar graph such that the edges incident to every $v \in V(G)$ are numbered $1, ..., D$. Formally there is a function $N: V \times [D] \rar V$ such that $N(v, i) = w$ iff $w$ is the $i^{th}$ neighbor of $v$. 

Then a rotation map is a function $Rot: V \times [D] \rar V \times [D]$ such that for all $v, w \in V$ and $i, j \in [D]$, $Rot(v, i) = (w, j)$ iff the $i^{th}$ neighbor of $v$ is $w$ and the $j^{th}$ neighbor of $w$ is $v$. 
\end{defn}

For technical reasons, we need a special kind of rotation map called a local inversion function. This is a rotation map where if $(v, i)$ maps to $(w, j)$ then $j$ only depends on $i$. 

\begin{defn}(Local inversion function)
Let $G$ be a $D$-regular graph with a rotation map $Rot: V \times [D] \rar V \times [D]$. A local inversion function $\phi_G: [D] \rar [D]$ is a permutation on $[D]$ such that for all $v \in V, i \in [D]$, 

$$Rot(v, i) = (N(v, i), \phi_G(i))$$
\end{defn}

We are ready to define the wide replacement product walk. Instead of the usual inner graph $H$ we use a ``wide'' inner graph on $D_1^s$ vertices for some integer $s \geq 1$. The vertices of $H$ correspond to $s$-tuples that define $s$ local inversion functions. The walk cycles through them. 

To take a step in the usual replacement product walk, we start at some vertex $v \in G \rrp H$ 
then compose two steps: an intra-cloud step which changes the $H$-component, and an inter-cloud step which changes the $G$-component. Every vertex in $G \rrp H$ is incident to a unique inter-cloud edge; therefore, there is only one choice of neighboring cloud, and so the position after the intra-cloud step determines the entire step. 

The $s$-wide replacement walk modifies the inter-cloud step so that there are $s$ choices during inter-cloud step. If $G$ is $D_1$-regular, then a vertex of $H$ corresponds to some vector $(a_0, ..., a_{s-1}) \in [D_1]^s$. The wide replacement walk maintains a clock which tracks how many steps have been taken. At time step $t$, the clock is set to $\ell = t \mod s$, and the inter-cloud step moves to a neighboring cloud according to the value of $a_\ell \in [D_1]$. 

After deciding which neighboring cloud to move to, the choice of which vertex in the cloud to land in is also determined by $a_\ell$. The walk updates the $H$-component by feeding the $\ell^{th}$ coordinate to the local inversion function $\phi_G: [D_1] \rar [D_1]$ of $G$, and leaving all other coordinates unchanged. So $(a_0, ..., a_{s-1}) \in [D_1]^s$ is mapped to $(a_0, ..., a_{\ell-1}, \phi_G(a_\ell), a_{\ell + 1}, ..., a_{s-1})$. This completes the inter-cloud step. 

The utility of the wide replacement walk is that the $H$-component of a vertex now stores $O(s \log(D_1))$ bits of information, rather than just $O(\log(D_1))$ bits. As we discussed in the introduction, the barrier to bias amplification is when the walk distribution is uniform within clouds. 

Now, the values of the $H$-component are precisely the instructions for the inter-cloud steps of the walk; therefore, the fact that the $H$-component is uniform is no longer bad news, since it means that the inter-cloud steps of the replacement walk imitate the truly random walk on the outer graph for the next $s$ steps. 



\begin{defn}
Let $G$ be a $D_1$-regular graph with local inversion function $\phi_G: [D_1] \rar [D_1]$. Let $H$ be a $D_2$-regular graph on $D_1^s$ vertices, for integer $s \geq 1$. A random step in the wide replacement product is determined as follows. 

Let ${(v^{(1)}, v^{(2)}) \in V(G) \times V(H)}$ be the current state of the walk at time $t \in \NN$. Sample random $i \in [D_2]$. Then the time-$t$ step according to $i$, denoted $Step_{i, t}(v^{(1)}, v^{(2)})$ is given by the composition of two steps: 

(i) Intra-cloud step: Leave the $G$-component $v^{(1)}$ unchaged. Move the $v^{(2)}$ component to its $i^{th}$ neighbor in $H$. Formally, set  
\begin{align}
w^{(1)} = v^{(1)} \\ w^{(2)} = v^{(2)}[i]
\end{align}

(ii) Inter-cloud step: Identifying $V(H)$ with $[D_1]^s$, let $\pi_j: [D_1]^s \rar [D_1]$ be projection onto the $j^{th}$ coordinate. Write $w^{(2)} \in V(H)$ as 
$w^{(2)} = (\pi_0(w^{(2)}), ..., \pi_{s-1}(w^{(2)})) \in [D_1]^s$. 

Let $\ell = t \mod s$. Move to the neighbor of $w^{(1)}$ in $G$ that is numbered by $\pi_\ell(w^{(2)}) \in D_1$. Then, update the $\ell^{th}$ coordinate of $H$-component $w^{(2)}$ by the local inversion function $\phi_G: [D_1] \rar [D_1]$ and leave other coordinates unchaged. Formally, let $\psi_{\ell}: [D_1]^s \rar [D_1]^s$ be 
$$\psi_\ell(a_0, ..., a_{s-1}) = (a_0, ..., a_{\ell-1}, \phi_G(a_\ell), a_{\ell+1}, ..., a_{s-1})$$

Set

\begin{align}
Step_{i, t}(v^{(1)}, v^{(2)}) = (w^{(1)}[\pi_\ell(w^{(2)})], \psi_{\ell}(w^{(2)}))
\end{align}
\end{defn}


A few remarks are in order. First, notice that the number of random bits needed to specify a random step is only $O(\log(D_2))$, despite the fact that we are moving on a graph with $V(G) \times V(H)$ vertices. This will be crucial in the analysis of the tradeoff between bias amplification and size increase of the small-bias set.  

Second, once a value of $t$ is fixed, so the clock is set to $\ell = t \mod s$, the wide replacement walk can be regarded as taking a usual step in the usual replacement walk. The intra-cloud step is unchaged, and the inter-cloud step depends only on the $\ell^{th}$ coordinate of the $H$-component. 


Since we have specified what it means to take a random step, this is sufficient to describe the walk. We simply initialize at a uniform vertex of $V(G) \times V(H)$ and then take some number of steps, to be chosen later. 

%% file: body.tex
 
\allowdisplaybreaks
\section{Expanding Generating Sets for Abelian Groups}

Throughout this section, let $G$ be a finite abelian group and $n \geq 1$. In this section, we will describe an efficient deterministic algorithm to construct a generating set $S \subset G^n$ such that the Cayley graph $Cay(G^n, S)$ has second eigenvalue at most $\eps$. The degree is $\abs{S} = O(\frac{n \log(\abs{G})^{O(1)}}{\eps^{2 + o(1)}})$. 

The inputs to our algorithm are a generating set $G^\prime \subset G$, integer $n \geq 1$, and desired expansion $\eps > 0$. The algorithm proceeds as follows: 

(i) Construct an $\eps_0$-biased set $S_0 \subset G^n$ with support size $O(n \log(\abs{G})^{O(1)})$ for a constant $\eps_0 < 1$. 

(ii) Perform a wide replacement walk to amplify the bias of $S_0$ to $\eps$. Specifically, we identify $S_0$ with the vertices of an outer graph $\Gamma$, and then choose an inner graph $H$ in a manner described later. We emphasize that while $\Gamma$ is an expander graph whose vertex set is $S_0$, it is not required to be a Cayley graph on $S_0$. For the purposes of this step, the group structure of $G$ is irrelevant. 

Let $t \geq 1$ be the walk length, to be chosen later. The output $\eps$-biased set $S \subset G^n$ corresponds to length-$t$ walks on the wide replacement product of $\Gamma$ and $H$. Given a sequence of vertices $(x_0, ..., x_t) \in V(\Gamma) \times V(H)$, we add up the components corresponding to $V(\Gamma)$, which are just elements of $S_0$, to obtain some element of $G^n$. This gives the elements of $S$. 

Next, let us informally describe parameter choices (precise choices are in section \ref{params-sec}). Let $D_2$ be the degree of $H$. At every step in the wide replacement walk we need to specify some $i \in [D_2]$ to take a step. It follows that $S \subset G^n$ has a size of $O(n \log(\abs{G})^{O(1)} \cdot D_2^{t})$. We must choose $t$ large enough to shrink the bias to $\epsilon$. The choice $t$ (walk length) and $D_2$ (degree of the inner graph) will determine the overall size of the output generating set. 


These choices hinge on the bias amplification bound of the wide replacement walk. We show that the $s$-wide replacement walk shrinks the bias by a factor of $O(s^2 \cdot \lambda(H)^{s-3})$ every $s$ steps. However, the size of the walk distribution grows by a factor of $O(D_2^{s})$ every $s$ steps. This imperfect bias amplification is why we cannot get optimal dependence on $\eps$, as that would require that the bias shrinks by exactly $O(\lambda(H)^{s})$ every $s$ steps. 

Therefore we cannot choose $H$ to be an optimal spectral expander with $\lambda(H) = \Theta(\frac{1}{\sqrt{D_2}})$. Instead, optimizing for the size of the output distribution, we set $s = \Theta(\frac{\log(1/\eps)^{1/3}}{\log\log(1/\eps)^{1/3}})$, second eigenvalue $\lambda(H) = \Theta( \frac{s \cdot \log(D_2)}{\sqrt{D_2}})$, and the walk length $t = \Theta(\frac{\log(1/\eps)}{\log(1/\lambda(H))} \cdot \frac{s^2}{s^2 - 5s + 1}) = \Theta((\frac{\log(1/\eps)}{\log(1/\lambda(H))})^{1 + o(1)})$. This is exactly the reason our output set has a dependence of $O(\frac{1}{\eps^{2+o(1)}})$ rather than exactly $O(\frac{1}{\eps^{2}})$, and the same is true for \cite{ta2017explicit}. 

This section is organized as follows. In section \ref{ordinary-walk-sec}, we describe how one can identify the elements $S_0$ with the vertices of an expander graph, and then perform the ordinary random walk on the graph to amplify the bias of $S_0$, albeit suboptimally. In section \ref{wide-replacement-walk-sec} we show how to express the bias of a wide replacement walk algebraically. In section \ref{matrix-norm-bound-sec} we prove an upper bound on this algebraic expression, therefore proving the bias amplification bound of the wide replacement walk. Finally, in section \ref{params-sec} we describe the details and exact parameters for the wide replacement walk, as well as the $\eps_0$-biased subset of $G^n$. 

\subsection{The ordinary expander walk}\label{ordinary-walk-sec}

Let $G$ be a finite abelian group. For ease of notation, we will refer to $G$ rather than $G^n$ until section \ref{params-sec}, when we need to discuss parameters. Since $H^n$ is a finite abelian group for all abelian $H$, there is no loss of generality. 

In this section we will show how to amplify the bias of a small-bias set in $G$ by performing a random walk on an expander. This will be a lemma in the analysis of our actual construction, which involves a \textit{wide replacement walk}. 

To state the bias amplification theorem, we need some notation. 

Let $G = \ZZ_{d_1} \oplus \cdots \oplus \ZZ_{d_k}$ be the invariant factor decomposition of $G$. Notice that $d_i | d_j$ for any $i < j$. In particular, all $d_i$ divide $d_k$. For $x \in G$ write $x = (x_1, ..., x_k)$, so that $x_i \in \ZZ_{d_i}$ for each $i$. 

Fix a nontrivial character $\chi: G \rar \CC^*$ corresponding to a group element $a \in G$. Let $a = (a_1, ..., a_k)$. Then for a given $x \in G$, $\chi(g) = \omega_{d_1}^{a_1 \cdot x} \cdots \omega_{d_k}^{a_k \cdot x}$. Since all $d_i$ divide $d_k$, we can write this as 

$$\chi(g) = \omega_{d_k}^{\sum_{i=1}^{k} (\frac{d_k}{d_i} a_i \cdot x_i) \mod d_k}.$$

Now, let $S_{init} \subset G$ have bias $\eps_0$. Identify $S_{init}$ with the vertices of some degree-regular expander graph $\Gamma$. We write $V \defeq V(\Gamma) = S_{init}$. In order to understand the bias of a random walk on $\Gamma$ with respect to $\chi$, we have to track how often the walk enters vertices which map to $\omega_{d_k}, \omega_{d_k}^2$, and so on. 

We will partition $S_{init}$ as follows. For $y \in \ZZ_{d_k}$, let $S_y$ be the elements of $S_{init}$ which are mapped to $\omega_{d_k}^{y}$ by $\chi$. Formally, $S_y = \{x \in S_{init}: y = (\sum_{i=1}^k \frac{d_k}{d_i} x_i \cdot a_i) \mod d_k \}$. Observe that $\{S_y: y \in \ZZ_{d_k}\}$ is a partition of $S_{init}$. 

Next, let $t > 0$ be the walk length. We will partition all length-$(t+1)$ sequences in $S_{init}$ according to their sum. For $y \in \ZZ_{d_k}$, let $T_y = \{b \in \ZZ_{d_k}^{t+1}: (\sum_i b_i) \mod d_k = y\}$. Again, notice that $\{T_y: y \in \ZZ_{d_k}\}$ is a partition of $\ZZ_{d_k}^{t+1}$. 

Finally, fix $y \in \ZZ_{d_k}$. The set $S_y$ corresponds to some subset of the vertices of $\Gamma$. Therefore we can identify $S_y$ with an $\abs{S_y}$-dimensional subspace of $\CC^{V}$. Let $\Pi_{y}: \CC^{V} \rar \CC^{V}$ be the projection matrix onto this subspace. Let $\Pi = \sum_{y \in \ZZ_{d_k}} \omega_{d_k}^y \Pi_y$. We write $\Pi = \Pi(\chi)$ to indicate the dependence on choice of $\chi$. 

We can now state the bias amplification theorem for ordinary expander walks. 

\begin{thrm}[Ordinary $t$-step expander walk]
Let $S_{init} \subset G$ have bias $\eps_0$ and let $\Gamma = (S_{init}, E)$ be a $d$-regular expander graph with $\lambda(\Gamma) = \lambda < 1$. Suppose $D \sim G$ is the distribution induced by beginning at a uniform vertex and taking a $t$-step random walk $(x^{(0)}, ..., x^{(t)})$  and then adding the results of the walk to get an element $(\sum_i x^{(i)}) \in G$. 

Let $\chi^*: G \rar \CC^*$ be the nontrivial character which maximizes the bias of $D$. Let $\Pi = \Pi(\chi^*)$, and $\norm \cdot \norm$ be the matrix operator norm. Finally, abusing notation, let $\Gamma$ be the random walk matrix of $\Gamma$. Then, 

$$bias(D) = bias(\chi^*) \leq \norm (\Pi \Gamma)^t \Pi \norm$$

\end{thrm}

\begin{proof}
Let $u = \frac{1}{\sqrt{\absl V(\Gamma) \absr}} \vec{1}$ be the normalized all-ones vector. Let $a^* \in G$ be the element corresponding to $\chi^*$. Let $(a_1^*, ..., a_k^*) \in \ZZ_{d_1} \oplus \cdots \oplus \ZZ_{d_k}$ denote $a^*$ written in the invariant factor decomposition. 

Let $W \sim V^{t+1}$ denote the distribution of all $t$-step walks on $\Gamma$. Let $(x^{(0)}, ..., x^{(t)}) \sim W$ be some sequence of random walk steps. So $x^{(0)} \sim S_{init}$ (since the walk begins at a uniformly random vertex) $x^{(i+1)}$ is a uniformly random neighbor of $x^{(i)}$. If $\vec{v}^{(i)} \in \CC^V$ is the distribution at step $i$, then $\vec{v}^{(i+1)} = \Gamma \vec{v}^{(i)}$. 

Recall that we use subscripts to denote invariant factors, so $x = (x_1, ..., x_k) \in \bigoplus\limits_{i=1}^{k} \ZZ_{d_i}$. 

\begin{eqnarray}
Bias(D) &=& Bias_D(\chi^{*}) \\
&=& \absl \EE_{(x^{(0)}, ..., x^{(t)}) \sim W} \prod\limits_{i=1}^{k} \omega_{d_i}^{x_i \cdot a_i^*} \absr \\
&=& \absl \EE_{(x^{(0)}, ..., x^{(t)}) \sim W} \omega_{d_k}^{\sum\limits_{i=1}^{k} \frac{d_k}{d_i} x_i \cdot a_i^*} \absr \\
&=& \absl \sum_{y \in \ZZ_{d_k}} \omega_{d_k}^y \PP_{(x^{(0)}, ..., x^{(t)}) \sim W}[y = (\sum_{j=0}^{t} \sum_{i=1}^{k} \frac{d_k}{d_i} x_i^{(j)}\cdot  a_i^*) \mod d_k] \absr \\
&=& \absl \sum_{y \in \ZZ_{d_k}} \sum\limits_{b \in T_y} \omega_{d_k}^y \PP_{(x^{(0)}, ..., x^{(t)}) \sim W}[\bigwedge\limits_{j=0}^{t} (x^{(j)} \in S_{b_j})] \absr \\
&=& \absl \sum_{y \in \ZZ_{d_k}} \omega_{d_k}^y (u^T \sum_{b \in T_y} \Pi_{b_t} \Gamma \cdots \Pi_{b_1} \Gamma \Pi_{b_0} u) \absr \\
&=& \absl u^T (\sum_{b \in \ZZ_{d_k}^{t+1}} \omega_{d_k}^{\sum_j b_j} \Pi_{b_t} \Gamma \cdots \Pi_{b_1} \Gamma \Pi_{b_0}) u \absr \\
&=& \absl u^T (\sum_{b_t \in \ZZ_{d_k}} \omega_{d_k}^{b_t} \Pi_{b_t}) \Gamma \cdots (\sum_{b_1 \in \ZZ_{d_k}} \omega_{d_k}^{b_1} \Pi_{b_1}) \Gamma (\sum_{b_0 \in \ZZ_{d_k}} \omega_{d_k}^{b_0} \Pi_{b_0}) u \absr \\
&=& \absl u^T (\Pi \Gamma)^t \Pi u \absr \\
&\leq& \norm (\Pi \Gamma)^t \Pi \norm 
\end{eqnarray} \end{proof}

We have thus obtained an algebraic expression for the bias of the walk distribution, which we will now upper-bound. 

\begin{thrm}[Matrix norm bounds]\label{thrm-matrix-norm}
Let $\Pi, \Gamma$ be as before. 

(i) $\norm \Pi \norm = 1$. 

(ii) $\norm (\Pi \Gamma)^2 \norm \leq \eps_0 + 2 \lambda$

It follows that $\norm (\Pi \Gamma)^t \Pi \norm \leq (\eps_0 + 2\lambda)^{\floor*{t/2}}$. 
\end{thrm}

\begin{proof}
(i) Fix $v \in \CC^S$. Then let $\Pi_{i, i} = \omega_d^{k_i}$ for some integers $d, k_i$. Then

$$\norm \Pi v \norm = (\sum_i \absl \omega_d^{k_i} v_i \absr^2)^{1/2}) = (\sum_i \absl \omega_d^{k_i} \absr^2 \absl v_i \absr^2)^{1/2} = (\sum_i \absl v_i \absr^2)^{1/2} = \norm v \norm$$

Restricting to unit vectors $v$, it follows that $\norm \Pi \norm = 1$. 

(ii) Let $v \in \CC^S$ be a unit vector. Let $v^{\pll}$ be its projection onto $\la \vec{1} \rangle$ (the one-dimensional subspace of parallel vectors), and let $v^\perp$ be its projection onto the orthogonal complement $\la \vec{1} \ra^\perp$. Then $v = v^\pll + v^\perp$. 

Let $u = \frac{1}{\sqrt{\absl S \absr}} \vec{1}$ be the normalized all-ones vector. Let $\lambda \defeq \lambda(\Gamma)$. Then observe that

\begin{align}
\norm (\Pi \Gamma)^2 v \norm \leq \norm (\Pi \Gamma)^2 v^\pll \norm + \norm (\Pi \Gamma)^2 v^\perp \norm \\
\leq \norm v^\pll \norm \norm (\Pi \Gamma)^2 u \norm + \norm \Pi \Gamma \Pi \norm \norm \Gamma v^\perp \norm \\
\leq \norm \Pi \Gamma \Pi u \norm + \lambda \\
\leq \norm \Pi \Gamma (\Pi u)^\pll \norm + \norm \Pi \Gamma (\Pi u)^\perp \norm + \lambda \\
\leq \norm \Pi (\Pi u)^\pll \norm + 2\lambda \\
\leq \norm \Pi \norm \norm (\Pi u)^\pll \norm + 2\lambda \\
\leq \norm (\Pi u)^\pll \norm + 2\lambda 
\end{align}

It remains to show that $\norm (\Pi u)^\pll \norm \leq \eps_0$. Observe that 
\begin{align}
\norm (\Pi u)^\pll \norm =\absl \la u, \Pi u \ra \absr \norm u \norm  \\
= \absl \la u, \Pi u \ra \absr \\
= \frac{1}{\absl S_{init} \absr} \absl \sum_{i=1}^{\absl S_{init} \absr} \Pi_{i, i} \absr \\
= \frac{1}{\absl S_{init} \absr} \absl \sum_{y \in \ZZ_d} \omega_d^y \absl S_y \absr \absr \\
= \absl \sum_{y \in \ZZ_d} \omega_d^y \PP_{x \sim S_{init}}[\chi_{a^*}(x) = \omega_{d_k}^y] \absr \\
= \absl bias_{S_{init}}(\chi_{a^*}) \absr \\
\leq \eps_0
\end{align}

We conclude that $\norm (\Pi \Gamma)^2 v \norm \leq \eps_0 + 2\lambda$. 
\end{proof}

Combining the two propositions in this section, it follows that a $t$-step walk amplifies the bias to $(\eps_0 + 2\lambda)^{\floor{t/2}}$. 

\subsection{The wide replacement walk}\label{wide-replacement-walk-sec}

In this section and the subsequent one, we will show how the wide replacement walk amplifies bias more efficiently than an ordinary expander walk. We will proceed in a similar manner to the last section, by first obtaining an algebraic expression for the bias of the random walk distribution, and then upper-bounding the algebraic expression in section \ref{matrix-norm-bound-sec}. 

\subsubsection{Setup}

Let $\Gamma = (S_{init}, E)$ be a graph whose vertices are some constant-bias set $S_{init} \subset G$ as before. Suppose $\Gamma$ is $D_1$-regular. Let $\phi_\Gamma: [D_1] \rar [D_1]$ be the local inversion function of $\Gamma$. 

Let $s > 0$ be an integer, and let $H$ be a $D_2$-regular expander graph on $[D_1]^s$ vertices. We will abuse notation and use $\Gamma, H$ to denote the random walk matrices of $\Gamma, H$ respectively.

Let $V^1 = \CC^{S_{init}} = \CC^{V(\Gamma)}$ and $V^2 = \CC^{D_1^s} = \CC^{V(H)}$. We define three operators on $V^1 \otimes V^2$ that we need to describe the bias of the wide replacement walk. Let $v^1 \otimes v^2 \in V^1 \otimes V^2$. 

For $i \in [s]$ define the projection matrix $P_i: V^2 \rar \CC^{D_1}$ as follows. Notice $V^2 = \CC^{V(H)} \cong \CC^{D_1^s}$. Identifying $V(H)$ with $\ZZ_{D_1}^s$, let $Z_i \subset V(H)$ correspond to $\{(0, ..., 0, a_i, 0, ..., 0) \in \ZZ_{D_1}^s: a_i \in \ZZ_{D_1}\}$. So we can identify $Z_i \subset V(H)$ with a $D_1$-dimensional subspace of $\CC^{V(H)}$. Then let $P_i: V^2 \rar \CC^{D_1}$ be the projection onto this subspace. 

Given some $v^1 \in V^1$ and $j \in [D_1]$, the vector $v^1[j] \in V^1$ is a permutation of the coordinates of $v^1$ based on the mapping of each vertex to its $j^{th}$ neighbor in $\Gamma$ \footnote{This is well-defined as long as the graph $\Gamma$ is $d$-regular, since its adjacency matrix is then just a sum of $d$ permutation matrices.}. This corresponds to taking a step in $\Gamma$, by moving along the edge numbered $j$ incident to the current vertex. For $w \in \CC^{D_1}$, let $v^1[w] = \sum_{j=1}^{D_1} w_j \cdot v_1[j]$. 

Finally, given the local inversion function $\phi_{\Gamma}: [D_1] \rar [D_1]$ of $\Gamma$ and $i \in [s]$, define $\psi_{\Gamma}^{(i)}: [D_1]^s \rar [D_1]^s$ as the function which applies $\phi_\Gamma$ to the $i^{th}$ coordinate and leaves other coordinates unchanged. Since $\phi_\Gamma$ is a permutation on $[D_1]$, $\psi_{\Gamma}^{(i)}$ is a permutation on $[D_1]^s$. Abusing notation, let $\psi_{\Gamma}^{(i)}: \CC^{D_1^s} \rar \CC^{D_1^s}$ denote the permutation matrix which permutes coordinates according to $\psi_{\Gamma}^{(i)}$. 

We are ready to define the three operators which describe the bias of the wide replacement walk. 

\begin{align}
\HH(v^1 \otimes v^2) &= v^1 \otimes H(v^2) \\
\forall \chi \in \hat{G}, y \in \ZZ_d: \PPi_y(\chi)(v^1 \otimes v^2) &= \Pi_y(\chi)(v^1) \otimes v^2 \\
\begin{split}
\forall \ell \in \{0, 1, ..., s-1\}: \Ga_\ell(v^1 \otimes v^2)
&= v^1[P_\ell(v^2)] \otimes \psi_{\Gamma}^{(\ell)}(v^2)
\end{split}
\end{align}

Note that each of these operators is a tensor product of operators on $V^1, V^2$, and hence preserves tensor products. 


Moreover, notice $\HH, \Ga_{t \mod s}$ are precisely the transition matrices of the $H$-step and $\Gamma$-step in the wide replacement walk at time $t$.  

For a character $\chi: G \rar \CC^*$ let $\PPi(\chi) = \sum_{y \in \ZZ_{d_k}} \omega_{d_k}^y \PPi_y(\chi)$. $\PPi$ plays the role of $\Pi$ from the analysis of the ordinary expander walk. 

\newcommand{\LL}{\dot L}

For notational convenience,

$$\LL_j(\chi) \defeq \PPi(\chi) \Ga_j \HH$$

\subsubsection{Algebraic Expression for the Bias}

\allowdisplaybreaks

In this section we will express the bias of the wide replacement walk distribution in terms of the matrix norms of $\LL_0, ..., \LL_{s-1}$. 

\begin{prop}[$t$-step $s$-wide replacement product walk]
Let $G$ be a finite abelian group. Let $S_{init} \subset G$ have bias $\eps_0$ and let $\Gamma = (S_{init}, E)$ be a $D_1$-regular expander graph. Let $H$ be a $D_2$ regular expander on $[D_1]^s$ vertices for some integer $s \geq 1$. 

Let $D_{walk} \sim G$ be the $t$-step $s$-wide replacement product walk distribution. It is defined by beginning at a uniform vertex and performing an $t$-step wide replacement wide on $V(\Gamma) \times V(H)$. Given a sequence of vertices $((a_0, b_0), ..., (a_t, b_t)) \in V(\Gamma) \times V(H)$ obtained from a walk, we output $(\sum_i a_i) \in G$. Then $D_{walk} \sim G$ is the distribution induced by taking all such $t$-step walks. 

We claim that if $\chi^*: G \rar \CC^*$ is the nontrivial character which maximizes the bias of $D_{walk}$, and $\PPi = \PPi(\chi^*)$, then using the notation from above, 

$$bias(D_{walk}) = bias(D_{walk}, \chi^*) \leq \norm \LL_{s-1}(\chi^*) \cdots \LL_{0}(\chi^*) \norm^{\floor{t/s}}$$
\end{prop}

\begin{proof}
First, we recall the notation defined in section \ref{ordinary-walk-sec}. 

Let $a^* \in G$ be the element corresponding to the the character $\chi^*$. Recall that $G \cong \ZZ_{d_1} \oplus \cdots \oplus \ZZ_{d_k}$ and we write $z \in G$ as $z = (z_1, ..., z_k)$, where $z_i \in \ZZ_{d_i}$. In particular $a^* = (a_1^*, ..., a_k^*)$ where each $a_i^* \in \ZZ_{d_i}$. 

For $y \in \ZZ_{d_k}$ let $S_y \subset S_{init}$ be the elements mapped to $\omega_{d_k}^{y}$ by $\chi^*$. Let $T_y \subset \ZZ_{d_k}^{t+1}$ be sequences which sum to $y$. 

Let $W \sim (V(\Gamma) \times V(H))^{t+1}$ be the distribution of all length-$t$ wide replacement walks, starting at a uniform vertex. 

Next, let $u^1 \in V^1, u^2 \in V^2$ be the all-ones vectors scaled to be unit vectors in the $2$-norm. 
Let $u = u^1 \otimes u^2$.

Let $v^{(0)}, ..., v^{(t)} \in V^1 \otimes V^2$ be the distribution of steps corresponding to $W$. 
Note that $v^{(0)}$ is the all-ones vector scaled by $\frac{1}{\abs{V(\Gamma) \times V(H)}}$ 
(since the walk begins at a uniformly random vertex), 
and $v^{(i+1)}$ proceeds by taking a step in the wide replacement product graph from the distribution $v^{(i)}$. 
In particular, $v^{(i+1)} \in V^1 \otimes V^2$, since both $\Ga, \HH$ tensorize and the walk begins at $v^{(0)} = (u^1 \otimes u^2) \in V^1 \otimes V^2$. 

For any $v \in V^1 \otimes V^2$, let $v^1, v^2$ denote its $V^1, V^2$ components respectively, so that $v = v^1 \otimes v^2$. Similarly for $(a, b) \in V(\Gamma) \times V(H)$ let $(a, b)^{1} = a$ and $(a, b)^{2} = b$. 

The output of the random walk for some sample $x^{(0)}, ..., x^{(t)} \sim W$ is then $x = \sum_i (x^{(i)})^1 \in G$. Note that since $x \in G$, we can write $x$ in invariant factor form as $x = (x_1, ..., x_k)$, where $x_i \in \ZZ_{d_i}$. 


The bias of $D_{walk}$ is thus: 

\begin{eqnarray}
Bias(D_{walk}) &=& Bias(D_{walk}, \chi^*) \\
&=& \absl \EE_{x^{(0)}, ..., x^{(t)} \sim W} \prod\limits_{i=1}^{k} \omega_{d_i}^{x_i \cdot a_i^*} \absr \\
&=& \absl \EE_{x^{(0)}, ..., x^{(t)} \sim W} [\omega_{d_k}^{\sum\limits_{i=1}^{k} \frac{d_k}{d_i} x_i \cdot a_i^*}] \absr \\
&=& \absl \sum_{y \in \ZZ_{d_k}} \omega_{d_k}^y \PP_{x^{(0)}, ..., x^{(t)} \sim W}[y = (\sum_{j=0}^{t} \sum_{i=1}^{k} \frac{d_k}{d_i} (x^{(j)})_i^1 \cdot a_i^*) \mod d_k] \absr \\ 
&=& \absl \sum_{y \in \ZZ_{d_k}} \sum_{b \in T_y} \omega_{d_k}^y \PP_{x^{(0)}, ..., x^{(t)} \sim W}[\bigwedge\limits_{j=0}^{t} (x^{(j)})^1 \in S_{b_j}] \absr \\
&=& \absl \sum_{y \in \ZZ_{d_k}} \omega_{d_k}^y (u^T \sum_{b \in T_y} \PPi_{b_t} \Ga_{t \mod s} \HH \cdots \PPi_{b_1} \Ga_{0} \HH \PPi_{b_0} u) \absr \\
&=& \absl u^T (\sum_{y \in \ZZ_{d_k}} \omega_{d_k}^y \sum_{b \in T_y} \PPi_{b_t} \Ga_{t \mod s} \HH \cdots \PPi_{b_1} \Ga_{0} \HH \PPi_{b_0}) u \absr \\
&=& \absl u^T \sum_{b \in \ZZ_{d_k}^{t+1}} \omega_{d_k}^{\sum b_i} \PPi_{b_t} \Ga_{t \mod s} \HH \cdots \PPi_{b_1} \Ga_{0} \HH \PPi_{b_0}) u \absr \\
&=& \absl u^T (\sum_{b_t \in \ZZ_{d_k}} \omega_{d_k}^{b_t} \PPi_{b_t} ) \Ga_{t \mod s} \HH \cdots (\sum_{b_1 \in \ZZ_{d_k}} \omega_{d_k}^{b_1} \PPi_{b_1}) \Ga_1 \HH (\sum_{b_0 \in \ZZ_{d_k}} \omega_{d_k}^{b_0} \PPi_{b_0}) u \absr \\
&=& \absl u^T \PPi \Ga_{t \mod s} \HH \cdots \PPi \Ga_1 \HH \PPi u \absr \\
&\leq& \norm \PPi \Ga_{t \mod s} \HH \cdots \PPi \Ga_1 \HH \PPi \norm \\
&\leq& \norm \PPi \Ga_{s-1} \HH \cdots \PPi \Ga_1 \HH \norm^{\floor{t/s}} \norm \PPi \norm \\
&\leq& \norm \LL_{s-1} \cdots \LL_{0} \norm^{\floor{t/s}} 
\end{eqnarray} 
\end{proof}

We have shown how to express the bias of the replacement walk distribution algebraically. It remains to be shown that this matrix norm is indeed bounded. To show that the wide replacement walk gains from $s - O(1)$ out of every $s$ steps, we need to show that $\norm \LL_{s-1} \cdots \LL_{0} \norm \leq \lambda(H)^{s - O(1)}$. We will do this in the next section. 

\subsection{Bounding the matrix norm}\label{matrix-norm-bound-sec}

In the previous section we showed that the bound the bias of the wide-replacement walk distribution, it suffices to bound the operator norm of the following matrix, defined with respect to the worst-case character $\chi^*$ of the walk distribution: $$\LL_{s-1} \cdots \LL_0$$

This is almost exactly the same matrix as the one analyzed in \cite{ta2017explicit}. The difference is that the operator $\PPi$, instead of tracking how often the walk enters the sets in a bipartition of $S_{init}$, now tracks how often the walk enters the sets in a $d_k$-way partition of $S_{init}$. Here $d_k = \Omega(\log(\abs{G}))$ is the largest invariant factor of $G$. 

As a consequence, the diagonal entries of $\PPi$ now come from the ${d_k}^{th}$ roots of unity, rather than $\{\pm 1\}$. The analysis of the matrix bound from \cite{ta2017explicit} mostly carries through, although working over $\CC^{V_1} \otimes \CC^{V^2}$ rather than the reals will require some care. 



As in \cite{ta2017explicit}, our argument will proceed by considering arbitrary vectors $v, w$ and analyazing $\la v, \LL_{s_1} \cdots \LL_0 w \ra$. We will repeatedly decompose the vectors into their parallel and perpendicular components. Let $V^\pll = V^1 \otimes \vec{1}$ denote vectors whose $H$-component is a scalar multiple of $\vec{1}$ (``parallel vectors''), and $V^\perp = (V^\pll)^{\bot}$ (``perpendicular vectors''). 

Because of the spectral expansion of $H$, every time a vector is in $V^\perp$ we can show it shrinks by a factor of $\lambda(H)$. The hard case is when vectors are in $V^\pll$. Here, we will prove a technical lemma which is a straightforward generalization of the core lemma in \cite{ta2017explicit}. The lemma shows if the walk distribution is in $V^\pll$, then any \textit{sequence} of $s$ steps imitates a random walk of $s$ steps on the outer graph $\Gamma$. This allows us to argue that the bias is amplified as though taking the ordinary random walk on $\Gamma$. If the bias so far is $\alpha$, then this scales the bias by $\alpha \mapsto (\alpha + 2\lambda(\Gamma))^{s/2}$ after $s$ steps. Notice that this step uses the bias amplification of the ordinary expander walk on $\Gamma$. 

This turns out to be enough. Let $\eps_0 = bias(S_{init})$ be the bias of the initial set $S_{init} \subset G$. Since $\eps_0$ is a constant, we can select graphs $\Gamma, H$ such that $\eps_0 + 2\lambda(\Gamma) \leq \lambda(H)^2$. Therefore, while we do not gain a factor of $(\lambda(\Gamma))^s$ every $s$ steps, we will gain according to a factor of $(\lambda(H))^{s - O(1)}$. Since $\lambda_2$ is bounded away from $1$, the difference between gaining according to $\lambda(\Gamma)$ or $\lambda(H)$ does not matter asymptotically. 

Therefore, whether in the $V^\perp$ or $V^\pll$ case, we shrink the bias by a factor of $\lambda(H)^{s - O(1)}$ for every $s$ steps. 

\subsubsection{Action on Parallel Vectors}

In this section we will prove that any sequence of $s$ steps imitates a random walk of $s$ steps on the outer graph $\Gamma$. This allows us to avoid the issue of getting stuck in ``parallel vectors'' in our overall parallel-perpendicular decomposition argument. 

We begin with an operator-averaging lemma. 

\begin{prop}[Operator-Averaging, \cite{ta2017explicit} Claim 14]
Let $\Omega$ be a finite set and $P, Q$ probability distributions on $\Omega$. Let $\norm P - Q \norm_1$ denote the difference of the distributions in the $1$-norm. Further, let $\{T_x\}_{x \in \Omega}$ be a family of linear operators on $\CC^n$ indexed by $\Omega$, such that for all $x \in\Omega$, $\norm T_x \norm \leq 1$. Let $A = \EE_{x \sim P}[T_x]$ and $B = \EE_{x \sim Q}[T_x]$. We claim that for all $v, w \in \CC^n$ that 

$$\absl \langle Av, w \rangle - \langle Bv, w \rangle \absr \leq \norm P - Q \norm_1 \norm v \norm \norm w \norm$$
\begin{proof}
\begin{align}
\absl \langle Av, w \rangle - \langle Bv, w \rangle \absr
&= \absl \sum_{x \in \Omega} (P(x) - Q(x)) \langle T_x v, w \rangle \absr \\
&\leq \sum_{x \in \Omega} \absl P(x) - Q(x) \absr \absl \langle T_x v, w \rangle \absr \\
&\leq \sum_{x \in \Omega} \absl P(x) - Q(x) \absr (\norm T_x \norm \norm v \norm \norm w \norm) \\
&\leq \norm v \norm \norm w \norm \norm P - Q \norm_1
\end{align}
\end{proof}
\end{prop}

Next, we generalize the key technical lemma of \cite{ta2017explicit}, which deals with the action of $\LL_i$ on $V^\pll$. We need to formalize the notion of the wide replacement walk ``imitating'' the ordinary random walk on the outer graph, which we do via the notion of a pseudorandom inner graph. 

\begin{defn}(Pseudorandom inner graph)
Let $\Gamma$ be a $D_1$-regular graph with local inversion function $\phi_\Gamma: [D_1] \rar [D_1]$. Let $H$ be a $D_2$-regular graph on $D_1^s$ vertices. Let $\zeta \geq 0$. We say $H$ is $\zeta$-pseudorandom with respect to $\Gamma$ if for all $s$-step sequences in the $s$-wide replacement walk, the corresponding $V^1$-instructions are $\zeta$-close to $Unif([D_1]^s)$ in $\ell_1$-norm. 

Formally, let the adjacency matrix of $H$ be $H = \frac{1}{D_2} \sum_{i=1}^{D_2} \Xi_i$, where each $\Xi_i$ is a permutation matrix \footnote{By the Birkhoff-von Neumann Theorem, the adjacency matrix of a $d$-regular graph is a sum of $d$ permutation matrices.}. Let $\xi_i: V(H) \rar V(H)$ be the permutation map corresponding to $\Xi_i$. For $0 \leq k < s$, let $\psi_k: [D_1]^s \rar [D_1]^s$ be $\psi_k(a_0, ..., a_{s-1}) = (a_0, ..., a_{k-1}, \phi_\Gamma(a_k), a_{k+1}, ..., a_{s-1})$. 

Fix $(j_0, ..., j_{s-1}) \in [D_2]^s$. For some $(u^1, u^2) \in V(\Gamma) \times V(H)$ let $\sigma_{j_0}(u^2) = \gamma_{j_0}(u^2)$. For $\ell > 0$, let 
$$\sigma_{j_\ell, ..., j_0}(u^2) = \gamma_{j_\ell}(\psi_{\ell-1}(\sigma_{j_{\ell-1}, ..., j_0}(u^2)))$$

We say $(j_0, ..., j_{s-1}) \in [D_2]^s$ is $\zeta$-pseudorandom with respect to $\Gamma$ if 

$$\norm (\pi_0(\sigma_{j_0}(Unif([D_1]))), ..., \pi_{s-1}(\sigma_{j_{s-1}, ..., j_0}(Unif([D_1])))) - Unif([D_1]^s) \norm_1 \leq \zeta$$

We say the inner graph $H$ is $\zeta$-pseudorandom with respect to the outer graph $\Gamma$ if for all $(j_0, ..., j_{s-1}) \in [D_2]^s$, $(j_0, ..., j_{s-1})$ is $\zeta$-pseudorandom with respect to $\Gamma$. 
\end{defn}

If we unravel the definition, this is simply requiring that $H$ is compatible with the edge labeling of $\Gamma$ in precisely the way that we want. Pseudorandomness is a strong condition on $H$ which, by definition, guarantees the wide-replacement walk imitates the ordinary walk on $\Gamma$ in a suitable sense. 

With this definition we can return to proving the lemma. We will begin by proving the pseudorandomness claim for the case where $D_2 = 1$. The case of general $D_2$ will follow by an application of operator averaging. 

\newcommand{\txi}{\Tilde{\Xi}}

\begin{prop}[Action on parallel vectors]
Let $\ell \leq s$. Suppose that the sequence $(j_0, ..., j_{\ell - 1}) \in [D_2]^s$ is $\zeta$-pseudoranom with respect to the local inversion function $\phi: [D_1] \rar [D_1]$. Let $\txi_{j_0}, ..., \txi_{j_{\ell-1}}$ denote the operators on $V^1 \otimes V^2$ corresponding to the permutations $\xi_{j_0}, ..., \xi_{j_{\ell-1}}$ on $V(H)$. Let $\mathrm{1}_{V(H)}$ denote the normalized all-ones vector of length $\absl V(H) \absr$. 

For any $\tau = \tau^{1} \otimes \mathrm{1}_{V(H)}$ and $\upsilon = \upsilon^{1} \otimes \mathrm{1}_{V(H)}$, 

$$\abs[\Big]{\la \PPi \Ga_{\ell-1} \txi_{j_{\ell-1}} \cdots \PPi \Ga_0 \txi_{j_0} \tau, \upsilon \ra - \la (\pi \Gamma)^\ell \tau^{1}, \upsilon^1 \ra} \leq \zeta \norm \tau \norm \norm \upsilon \norm$$
\end{prop}

\begin{proof}
First, we consider the action of an $\LL_i$ on a general product vector when $D_2 = 1$. Let $(u^1 \otimes u^2) \in V^1 \otimes V^2$. By the Birkhoff-von Neumann Theorem the random walk matrix $\Gamma$ is an average of $D_1$ permutation matrices. Let $\Gamma = \frac{1}{D_1} \sum_{r=1}^{D_1} \mathcal{G}_r$, where each $\mathcal{G}_r$ is a permutation matrix. 

At step $i$, the $\Gamma$-step in the replacement walk is chosen according to the $i^{th}$ coordinate of the $H$-vector. The $H$-vector itself is just the distribution a random step in $H$, starting from $u^2$. 

We need two new notations. First, let $\Xi_{j_i}: \CC^{V_2} \rar \CC^{V_2}$ denote the permutation matrix corresponding to $\xi_{j_i}$, so that $\txi_{j_i} = I \otimes \Xi_{j_i}$. 

Second, let $\psi_\Gamma^{(i)}$ be the operator which applies the local inversion function $\phi_\Gamma$ to the $i^{th}$ coordinate and leaves all else unchanged. 

Finally, recall $P_i: \CC^{D_1^s} \rar \CC^{D_1}$ is the projection onto the $D_1$-dimensional subspace corresponding to the indices $(a_0, ..., a_{s-1}) \in \ZZ_{D_1}^s$ where $a_j = 0$ for all $j \neq i$. Given some $z \in \CC^{D_1}$, let ${\mathcal{G}_{z} = \sum_{i=1}^{D_1} z_i \mathcal{G}_i}$. Then, 

\begin{align}
\PPi \Ga_{i} \txi_{j_{i}}(u^1 \otimes u^2) &= \PPi \Ga_i(u^1 \otimes \Xi_{j_i}(u^2)) \\
&= \PPi(\mathcal{G}_{P_i(\Xi_{j_i}(u^2))})(u^1) \otimes \psi_\Gamma^{(i)}(\Xi_{j_i}(u^2)) \\
&= \Pi \mathcal{G}_{P_i(\Xi_{j_i}(u^2))}(u^1) \otimes \psi_\Gamma^{(i)}(\Xi_{j_i}(u^2)) 
\end{align}

Notice that the $H$-step always evolves according to a composition of permutations, since each $\Xi_j$ is a permutation matrix, and each $\psi_\Gamma^{(i)}$ operator is also a permutation which simply applies the permutation $\phi_\Gamma$ to the $i^{th}$ coordinate and leaves the rest of the coordinates unchanged. Let us introduce the following notation: 

\begin{align}
\sigma_{j_0} &\defeq \xi_{j_0} \\
\sigma_{j_0, ..., j_i} &\defeq \xi_{j_i} \circ \psi_{i-1} \circ \xi_{j_0, ..., j_{i-1}}
\end{align}

Now, we can return to the main claim. Let $e_i \in \CC^{V(H)}$ be the $i^{th}$ standard basis vector. Observe that $\tau = \tau^1 \otimes \mathrm{1}_{V(H)} = \frac{1}{\sqrt{\absl V(H) \absr}} (\tau_1 \otimes \sum_{i = 1}^{\absl V_2 \absr} e_{i})$. Let $\pi_i: [D_1]^s \rar [D_1]$ be projection onto the $i^{th}$ coordinate. We can express the action of $M \defeq \PPi \Ga_{\ell-1} \txi_{j_{\ell-1}} \cdots \PPi \Ga_0 \txi_{j_0}$ on $\tau$ as 

\begin{eqnarray}
M \tau &=& M(\frac{1}{\sqrt{\absl V_2 \absr}} (\tau_1 \otimes \sum_{i = 1}^{\absl V_2 \absr} e_{i})) \\
&=& \frac{1}{\sqrt{\absl V_2 \absr}} \sum_{i = 1}^{\absl V_2 \absr} M(\tau_1 \otimes e_{i}) \\
&=& \frac{1}{\sqrt{\absl V_2 \absr}} \sum_{i = 1}^{\absl V_2 \absr} (\Pi \mathcal{G}_{\pi_{\ell-1}(\sigma_{j_0, ..., j_{\ell-1}}(i))} \cdots \Pi \mathcal{G}_{\pi_0(\xi_{j_0}(i))} \tau^1 \otimes (\psi_{\ell-1} \circ \Xi_{j_{\ell-1}} \circ \psi_{\ell-2} \cdots \circ \Xi_{j_0} e_i))
\end{eqnarray}

In the last step, notice that the composition of permutation operators $\psi_{\ell-1} \circ \Xi_{j_{\ell-1}} \circ \psi_{\ell-2} \cdots \circ \Xi_{j_0}$ is itself a permutation. Therefore there exists some permutation $\sigma$ on $V(H)$ such that $\psi_{\ell-1} \circ \Gamma_{j_{\ell-1}} \circ \psi_{\ell-2} \cdots \circ \Gamma_{j_0} e_i = e_{\sigma(i)}$. 

Next, we proceed by decomposing $\upsilon = \upsilon^1 \otimes \mathrm{1}_{V(H)} = \frac{1}{\sqrt{\absl V(H) \absr}} (\upsilon_1 \otimes \sum_{i = 1}^{\absl V(H) \absr} e_{i})$ in a similar fashion. 

\begin{align}
\la M \tau, \upsilon \ra - \la (\Pi \Gamma)^\ell \tau^{1}, \upsilon^1 \ra 
&= \la \frac{1}{\sqrt{\absl V_2 \absr}} \sum_{i = 1}^{\absl V_2 \absr} \Pi\mathcal{G}_{\pi_{\ell-1}(\sigma_{j_0, ..., j_{\ell-1}}(i))} \cdots \Pi \mathcal{G}_{\pi_0(\xi_{j_0}(i))} \tau^1 \otimes e_{\sigma(i)}, \upsilon^1 \otimes \sum_{j = 1}^{\absl V_2 \absr} e_{j} \ra \\
&= \frac{1}{\absl V_2 \absr} \sum_{i = 1}^{\absl V_2 \absr} \sum_{j = 1}^{\absl V_2 \absr} (\la \Pi \mathcal{G}_{\pi_{\ell-1}(\sigma_{j_0, ..., j_{\ell-1}}(i))} \cdots \Pi \mathcal{G}_{\pi_0(\xi_{j_0}(i))} \tau^1 \otimes e_{\sigma(i)}, \upsilon^1 \otimes e_j \ra) \\
&= \frac{1}{\absl V_2 \absr} \sum_{i = 1}^{\absl V_2 \absr} \sum_{j = 1}^{\absl V_2 \absr} 
(\la \Pi \mathcal{G}_{\pi_{\ell-1}(\sigma_{j_0, ..., j_{\ell-1}}(i))} \cdots \Pi \mathcal{G}_{\pi_0(\xi_{j_0}(i))} \tau^1, \upsilon^1\ra \nonumber \cdot \la e_{\sigma_i}, e_j \ra) \\
&= \frac{1}{\absl V_2 \absr} \sum_{i = 1}^{\absl V_2 \absr} \la \Pi \mathcal{G}_{\pi_{\ell-1}(\sigma_{j_0, ..., j_{\ell-1}}(i))} \cdots \Pi \mathcal{G}_{\pi_0(\xi_{j_0}(i))} \tau^1, \upsilon^1\ra
\end{align}

The last line follows from the fact $\la e_{\sigma(i)}, e_j \ra = 1$ if $\sigma(i) = j$ and $0$ otherwise. 

Next, we will use operator averaging. Define a distribution $Z$ on $[D_1]^{\ell}$ which corresponds to instructions for $\Gamma$-steps in the wide replacement walk. Formally, it picks some $i \in [V(H)]$ uniformly at random, and outputs $(z_0, ..., z_{\ell-1}) = (\pi_0(\sigma_{j_0, ..., j_{\ell-1}}(i))), ..., \pi_{\ell-1}(\sigma_{j_0, ..., j_{\ell-1}}(i)))$. 

By assumption, since $(j_0, ..., j_{\ell-1})$ is $\zeta$-pseudorandom with respect to $\Gamma$, the distribution $Z$ is $\zeta$-close to the uniform distribution on $[D_1]^{\ell}$ in the $L^1$-norm. In other words, the inter-cloud steps taken in the replacement product walk are close to imitating an ordinary random walk on $\Gamma$. Explicitly, the expectations of each distribution are: 

\begin{align}
\EE_{(z_0, ..., z_{\ell-1}) \sim Z}[\la \Pi \mathcal{G}_{z_{\ell-1}} \cdots \Pi \mathcal{G}_{z_0} \tau^1, \upsilon^1 \ra] 
&= \frac{1}{\absl V_2 \absr} \sum_{i = 1}^{\absl V_2 \absr} \la \Pi \mathcal{G}_{\pi_{\ell-1}(\sigma_{j_0, ..., j_{\ell-1}}(i))} \cdots \Pi \mathcal{G}_{\pi_0(\xi_{j_0}(i))} \tau^1, \upsilon^1\ra \\
\EE_{(z_0, ..., z_{\ell-1}) \sim \textrm{Unif}([D_1]^\ell)}[\la \Pi \mathcal{G}_{z_{\ell-1}} \cdots \Pi \mathcal{G}_{z_0} \tau^1, \upsilon^1 \ra]  
&= \la (\Pi G)^\ell \tau^1, \xi^1 \ra
\end{align}

Notice that $\norm \Pi \norm \leq 1$, since it is just a sum of disjoint projections scaled by roots of unity. Moreover, $\norm \mathcal{G}_i \norm \leq 1$ for all $i$, since $\mathcal{G}_i$ is simply a permutation matrix. It follows that $\norm \Pi \mathcal{G}_{z_{\ell-1}} \cdots \Pi \mathcal{G}_{z_0} \norm \leq 1$ for all $(z_0, ..., z_{\ell-1})$. 

Therefore, we conclude by operator averaging that

\begin{align}
\abs[\Big]{\la \PPi \Ga_{\ell-1} \txi_{j_{\ell-1}} \cdots \PPi \Ga_0 \txi_{j_0} \tau, \upsilon \ra - \la (\Pi \Gamma)^\ell \tau^{1}, \upsilon^1 \ra} 
&\leq \norm Z - \textrm{Unif}([D_1]^\ell)\norm_1 \cdot \norm \tau^1 \norm \cdot \norm \xi^1 \norm \\
&\leq \zeta \cdot \norm \tau \norm \cdot \norm \upsilon \norm 
\end{align}

\end{proof}

Next, proving the general case where $D_2 \neq 1$ follows from another application of operator averaging. In particular, the matrix $H$ is the average of $D_2$ permutation matrices. We will average over the choice of permutation in $H$. 

\begin{cor}[Generalized action on parallel vectors (\cite{ta2017explicit} Theorem 27)]
Suppose that $H$ is $\zeta$-pseudorandom with respect to the local inversion function $\phi_\Gamma$ of $\Gamma$. For every $i_1, i_2 \in \{0, 1, ..., s-1\}$, and every $\tau, \upsilon \in V^\pll$, 

$$\abs[\Big]{\la \LL_{i_2} \cdots \LL_{i_1} \tau, \upsilon \ra - \la (\Pi \Gamma)^{i_2 - i_1 + 1} \tau^{1}, \upsilon^1 \ra} \leq \zeta \norm \tau \norm \norm \upsilon \norm$$
\end{cor}

\begin{proof}
Let $\ell = i_2 - i_1 + 1$. There is a collection of $D_2$ permutation matrices $\Xi_1, ..., \Xi_{D_2}$ such that $H = \frac{1}{D_2} \suml_{j=1}^{D_2} \Xi_j$. Therefore, 

$$\LL_{i_2} \cdots \LL_{i_1} = \EE_{j_1, ..., j_{\ell} \sim \textrm{Unif}([D_2])} [\PPi \Ga_{i_2} \txi_{j_\ell} \cdots \PPi \Ga_{i_1} \txi_{j_1}]$$

We showed above that for any choice of $j_1, ..., j_{\ell}$ our desired inequality is true. 

Next, let ${d(\tau^1, \upsilon^1) \defeq \abs[\Big]{\la \LL_{i_2} \cdots \LL_{i_1} \tau, \upsilon \ra - \la (\Pi \Gamma)^{i_2 - i_1 + 1} \tau^{1}, \upsilon^1 \ra}}$. We bound $d(\tau^1, \upsilon^1)$ as: 

\begin{align}
d(\tau^1, \upsilon^1) &= \abs[\Big]{\la \EE_{j_1, ..., j_{\ell} \sim \textrm{Unif}([D_2])} [\PPi \Ga_{i_2} \txi_{j_\ell} \cdots \PPi \Ga_{i_1} \txi_{j_1}] \tau, \upsilon \ra  - \la (\Pi \Gamma)^{i_2 - i_1 + 1} \tau^{1}, \upsilon^1 \ra} \\
&= \abs[\Big]{\EE_{j_1, ..., j_{\ell} \sim \textrm{Unif}([D_2])} [\la \PPi \Ga_{i_2} \txi_{j_\ell} \cdots \PPi \Ga_{i_1} \txi_{j_1} \tau, \upsilon \ra - \la (\Pi \Gamma)^{i_2 - i_1 + 1} \tau^{1}, \upsilon^1 \ra]} \\ 
&\leq \EE_{j_1, ..., j_{\ell} \sim \textrm{Unif}([D_2])} \abs[\Big]{\la \PPi \Ga_{i_2} \txi_{j_\ell} \cdots \PPi \Ga_{i_1} \txi_{j_1} \tau, \upsilon \ra - \la (\Pi \Gamma)^{i_2 - i_1 + 1} \tau^{1}, \upsilon^1 \ra} \\
&\leq \EE_{j_1, ..., j_{\ell} \sim \textrm{Unif}([D_2])} \zeta \cdot \norm \tau \norm \cdot \norm \upsilon \norm \\
&= \zeta \cdot \norm \tau \norm \cdot \norm \upsilon \norm
\end{align}

The penultimate step follows from Jensen's inequality. 
\end{proof}

\subsubsection{A bound for the algebraic expression of bias}

In the previous section we proved a technical lemma which allows us to circumvent the issue of getting stuck in parallel vectors. With this lemma, we are ready to prove our main theorem which bounds the matrix norm of $\LL_{s_1} \cdots \LL_0$. 

Our argument will proceed by considering the action of the matrix $\LL_{s_1} \cdots \LL_0$ on an arbitrary vector, 
and then repeatedly decomposing vectors into their $V^\pll$ and $V^\perp$ components. Because of the spectral expansion of $H$, every time a vector is in $V^\perp$ we can show it shrinks by a factor of $\lambda_2 = \lambda(H)$. 

The hard case is when vectors are in $V^\pll$. Here, we will use the technical lemma from the previous section to argue that any \textit{sequence} of $s$ steps imitates a random walk on the outer graph $\Gamma$. This allows us to argue that the bias is amplified as though taking the ordinary random walk on $\Gamma$. This scales the bias by $(\eps_0 + 2\lambda_1)^{s/2}$ at every $s$ steps. 

This turns out to be enough, as we can assume that $\eps_0 + 2\lambda_1 \leq \lambda_2^2$. Therefore, while we do not gain a factor of $(\lambda_1)^s$ every $s$ steps, we will gain according to a factor of $(\lambda_2)^s$. Since $\lambda_2 < 1$, the difference between gaining according to $\lambda_2$ or $\lambda_1$ does not matter asymptotically. Notice that the bias amplification of the ordinary expander walk turns out to be crucial for the wide replacement walk. 

Therefore, whether in the $V^\perp$ or $V^\pll$ case, we gain a factor of $\lambda_2^{s - O(1)}$ for every $s$ steps. 

\begin{thrm}[Bounding algebraic expression for bias]
Suppose that: 

i) $H$ is $\zeta$-pseudorandom with respect to $\phi_\Gamma$

ii) $\eps_0 + 2\lambda(\Gamma) \leq \lambda(H)^2$

Then we obtain the following bound for the bias of the walk after $s$ steps. 

$$\norm \LL_{s-1} \cdots \LL_0 \norm \leq \lambda(H)^s + s \lambda(H)^{s-1} + s^2(
\lambda(H)^{s-2} + \zeta)$$
\end{thrm}

\begin{proof}
For a (column) vector $v$, let $v^*$ denote its conjugate transpose. Let $w_s, x_0 \in \CC^{\abs{V_1} \cdot \abs{V_2}}$ be unit vectors which maximize the bilinear form $w_s^* \LL_{s-1} \cdots \LL_0 x_0$, so that 

$$\absl w_s^* \LL_{s-1} \cdots \LL_0 x_0 \absr = \norm \LL_{s-1} \cdots \LL_0 \norm$$

We will repeatedly apply the parallel-perpendicular decomposition from the left and right hand side of the expression. Let us define some additional notation. First, for any linear operator $M$ let $M^*$ denote its adjoint (e.g. its conjugate transpose). We define the following ``intermediate'' vectors which will arise in the decomposition. 

\begin{align}
x_i = \LL_{i-1} x_{i-1}^{\perp} \\
w_{s-i} = \begin{cases}
\HH (\Ga_{s-i} \PPi^* w_{s-i+1})^{\perp} & i \text{ odd} \\
\HH (\Ga_{s-i} \PPi w_{s-i+1})^{\perp} & i \text{ even}
\end{cases} \\
p_{s-i} = \begin{cases}
\HH (\Ga_{s-i} \PPi^* w_{s-i+1})^{\pll} & i \text{ odd} \\
\HH (\Ga_{s-i} \PPi w_{s-i+1})^{\pll} & i \text{ even}
\end{cases}
\end{align}

Next, we can decompose from the right side to obtain: 
\begin{align}
w_s^* \LL_{s-1} \cdots \LL_0 x_0 &= w_s^* \LL_{s-1} \cdots \LL_0 x_0^\pll + w_s^* \LL_{s-1} \cdots \LL_0 x_0^\perp \\
&= w_s^* \LL_{s-1} \cdots \LL_1 x_1 + w_s^* \LL_{s-1} \cdots \LL_0 x_0^\perp \\
&= w_s^* x_s + \sum\limits_{i=0}^{s-1} w_s^* \LL_{s-1} \cdots \LL_i x_i^\pll
\end{align}

Next, we decompose from the left side. Let $0 \leq i \leq s-1$. Then 

\begin{align}
w_s^* \LL_{s-1} \cdots \LL_i x_i^\pll &= (\LL_{s-1}^* w_s)^* \LL_{s-2} \cdots \LL_{i} x_i^\pll \\
&= ((\PPi \Ga_{s-1} \HH)^* w_s)^* \LL_{s-2} \cdots \LL_{i} x_i^\pll \\
&= (\HH^* \Ga_{s-1}^* \PPi^* w_s)^* \LL_{s-2} \cdots \LL_{i} x_i^\pll \\
&= (\HH \Ga_{s-1} \PPi^* w_s)^* \LL_{s-2} \cdots \LL_{i} x_i^\pll \\
&= (\HH (\Ga_{s-1} \PPi^* w_s)^\perp)^* \LL_{s-2} \cdots \LL_{i} x_i^\pll 
+ (\HH (\Ga_{s-1} \PPi^* w_s)^\pll)^* \LL_{s-2} \cdots \LL_{i} x_i^\pll \\
&= w_{s-1}^* \LL_{s-2} \cdots \LL_{i} x_i^\pll + p_{s-1}^* \LL_{s-2} \cdots \LL_{i} x_i^\pll
\end{align}

We can continue in this manner. Notice that since $1$ is odd that $w_{s-2} = \HH(\Ga_{s-1} \PPi w_{s-1})^\perp$, since $(\Pi^*)^* = \Pi$. Further, notice that since $w_i \in V^\perp$, that $w_i^* x_i^\pll = 0$. Therefore, we continue our decomposition to obtain 
\begin{align}
w_s^* \LL_{s-1} \cdots \LL_i x_i^\pll = p_i^* x_i^\pll + \sum_{j=i+1}^{s-1} p_{j}^* \LL_{j-1} \cdots \LL_{i} x_i^\pll 
\end{align}

Combining expressions, we obtain: 

\begin{align}
w_s^* \LL_{s-1} \cdots \LL_0 x_0 
&= w_s^* x_s + \sum\limits_{i=0}^{s-1} w_s^* \LL_{s-1} \cdots \LL_i x_i^\pll \\
&= w_s^* x_s + \sum\limits_{i=0}^{s-1} (p_i^* x_i^\pll + \sum_{j=i+1}^{s-1} p_{j}^* \LL_{j-1} \cdots \LL_{i} x_i^\pll)
\end{align}

Next, we bound each of the terms above in absolute value. We will separate the expression into three terms: 
\begin{align}
\abs[\Big]{w_s^* \LL_{s-1} \cdots \LL_0 x_0 } 
&\leq  \abs{w_s^* x_s} \\
&+ \abs[\Big]{\sum\limits_{i=0}^{s-1} p_i^* x_i^\pll} \\
&+ \abs[\Big]{\sum\limits_{i=0}^{s-1} \sum_{j=i+1}^{s-1} p_{j}^* \LL_{j-1} \cdots \LL_{i} x_i^\pll}
\end{align}

\textit{First term}: 

\begin{align}
\abs{w_s^* x_s} &\leq \norm w_s \norm \norm x_s \norm \\
&\leq \norm x_s \norm \\
&= \norm \LL_{s-1} x_{s-1}^\perp \norm \\
&= \norm \PPi \Ga_{s-1} \HH x_{s-1}^\perp \norm \\
&\leq \norm \PPi \Ga_{s-1}\norm \cdot \lambda(H) \norm x_{s-1}^\perp \norm \\
&\leq \lambda(H) \norm x_{s-1}^\perp \norm \\
&\leq \lambda(H)^{s} \norm x_0 \norm \\
&\leq \lambda(H)^s 
\end{align}

The above analysis of $ \norm x_s \norm$ does not depend on the value of $s$, so we also obtain a corollary of $\norm x_i \norm \leq \lambda(H)^i$. 

\textit{Second term}: Next, notice that for even values of $i \geq 0$, 

\begin{align}
\norm p_{s-i} \norm &= \norm \HH (\Ga_{s-i} \PPi w_{s-i+1})^{\pll} \norm \\
&\leq \norm \HH \norm \norm \Ga_{s-i} \PPi w_{s-i+1} \norm \\ 
&\leq \norm \Ga_{s-i} \norm \norm \PPi \norm \norm w_{s-i+1} \norm \\
&\leq \norm w_{s-i+1}\norm
\end{align}

The case for odd $i$ attains the same bound, since $\norm \PPi^*\norm \leq 1$. Next, 

\begin{align}
\norm w_{s-i} \norm &= \norm \HH (\Ga_{s-i} \PPi w_{s-i+1})^{\perp} \norm \\
&\leq \lambda(H) \norm (\Ga_{s-i} \PPi w_{s-i+1})^{\perp}\norm \\
&\leq \lambda(H) \norm \Ga_{s-i} \PPi \norm \norm w_{s-i+1} \norm \\
&\leq \lambda(H) \norm w_{s-i+1} \norm
\end{align}

Therefore, since $\norm w_s \norm \leq 1$ we obtain $\norm w_{s-i}\norm \leq \lambda(H)^i$. We are ready to bound the second term in the overall expression. 

\begin{align}
\abs[\Big]{\sum\limits_{i=0}^{s-1} p_i^* x_i^\pll} &\leq \sum\limits_{i=0}^{s-1} \norm p_i \norm \cdot \norm x_i^\pll \norm \\
&\leq \sum\limits_{i=0}^{s-1} \lambda(H)^i \norm p_i \norm \\
&\leq \sum\limits_{i=0}^{s-1} \lambda(H)^i \lambda(H)^{s-i-1} \\
&\leq s \lambda(H)^{s-1}
\end{align}

\textit{Third term}: Finally, the third term in the expression collects all of the ``leftover'' terms which could not be simplified through parallel-perpendicular decomposition. These are precisely the parallel components of the vectors obtained at each step of the decomposition. To bound this term we will use the technical lemmas about the action of the $\LL_i$ operators on parallel vectors, which in turn use the pseudorandom machinery from the wide replacement product. 

Consider some $(i, j)$ such that $0 \leq i < j \leq s-1$. We wish to bound $\abs[\Big]{p_{j}^* \LL_{j-1} \cdots \LL_{i} x_i^\pll}$. 

Then $p_j = p_j^{1} \otimes \mathrm{1}_{V_2}$ and $x_i^\pll = (x_i^\pll)^{1} \otimes \mathrm{1}_{V_2}$. Since $H$ is $\zeta$-pseudorandom with respect to $\phi_\Gamma$, it follows that 

$$\abs[\Big]{p_{j}^* \LL_{j-1} \cdots \LL_{i} x_i^\pll - p_j^1 (\Pi \Gamma)^{j-i}(x_i^\pll)^1} \leq \zeta \cdot \norm x_i^\pll \norm \cdot \norm p_j \norm$$

Therefore, up to an additive error factor, the action of the replacement walk on parallel vectors is the same as the action of a truly random walk on the outer graph $\Gamma$. We have already analyzed this walk, and as we argued, it amplifies bias by a factor of $(\eps_0 + 2\lambda(\Gamma))$ every two steps. Therefore, 
\begin{align}
\abs[\Big]{p_{j}^* \LL_{j-1} \cdots \LL_{i} x_i^\pll} 
&\leq \abs[\Big]{p_j^1 (\Pi \Gamma)^{j-i}(x_i^\pll)^1} + \zeta \cdot \norm x_i^\pll \norm \cdot \norm p_j \norm \\
&\leq (\norm (\Pi \Gamma)^{j-i}\norm \cdot \norm x_i^\pll \norm \cdot \norm p_j \norm) + \zeta \cdot \norm x_i^\pll \norm \cdot \norm p_j \norm \\
&\leq ((\eps_0 + 2\lambda(\Gamma))^{\floor{\frac{j-i}{2}}} + \zeta) \norm x_i^\pll \norm \cdot \norm p_j \norm \\
&\leq (\eps_0 + 2\lambda(\Gamma))^{\floor{\frac{j-i}{2}}} + \zeta) \lambda(H)^i \lambda(H)^{s-(j+1)} \\
&\leq (\lambda(H)^{j-i-1} + \zeta) \cdot \lambda(H)^{i + s - j - 1} \\
&\leq \lambda(H)^{s-2} + \zeta
\end{align}

We can therefore bound the third term by $\abs[\Big]{\sum\limits_{i=0}^{s-1} \sum_{j=i+1}^{s-1} p_{j}^* \LL_{j-1} \cdots \LL_{i} x_i^\pll} \leq s^2 \cdot (\lambda(H)^{s-2} + \zeta)$. 

Putting it all together, we conclude
\begin{align}
\norm \LL_{s-1} \cdots \LL_0 \norm 
&= \absl w_s^* \LL_{s-1} \cdots \LL_0 x_0 \absr \\
&\leq  \abs{w_s^* x_s} + \abs[\Big]{\sum\limits_{i=0}^{s-1} p_i^* x_i^\pll} 
+ \abs[\Big]{\sum\limits_{i=0}^{s-1} \sum_{j=i+1}^{s-1} p_{j}^* \LL_{j-1} \cdots \LL_{i} x_i^\pll} \\
&\leq \lambda(H)^s + s \lambda(H)^{s-1} + s^2(\lambda(H)^{s-2} + \zeta)
\end{align} 
\end{proof}

\subsection{Parameters of the Construction} \label{params-sec}

In this section we describe how to optimize parameters such that the wide replacement walk construction achieves our desired support size. Our construction and hence the parameters we choose are almost identical to those discussed in Section 5 of \cite{ta2017explicit}. 

The algorithm is given integer $n \geq 1$, desired second eigenvalue $\eps > 0$, and an arbitrary generating set for a group $G$. 

It first generates an $\eps_0$-biased set $S_{init} \subset G^n$ of size $O(\frac{n\log(\abs{G})^{O(1)}}{poly(\eps_0)})$ for a constant $\eps_0$. For concreteness we set $\eps_0 = 0.1$. 

\begin{prop}
There exists a deterministic, polynomial time algorithm which, given a generating set for an abelian group $G$ and integer $n\geq 1$, outputs a generating set $S_{init} \subset G^n$ of size $O(n (\log(\abs{G}))^{O(1)})$ such that the Cayley graph has second eigenvalue at most $0.1$. 
\end{prop}

\begin{proof}
First, by Theorem 4 of \cite{chen-moore-russell}, we can construct a generating set $S \subset G$ with second eigenvalue $(1 - \frac{C}{\log\log(\abs{G})} + \beta)$ for a parameter $\beta$ and universal constant $C$. Its size will be $\abs{S} = O(\frac{n \log(\abs{G})}{\beta^{O(1)}}) = O(n\log(\abs{G})^2)$. Setting $\beta = \frac{C}{2 \log\log(\abs{G})}$, we obtain second eigenvalue $(1 - \frac{C}{2 \log\log(\abs{G})})$. 

Next, we can amplify the bias of $S$ to $0.1$ by taking a $t$-step ordinary expander walk. By the results of section $3.1$, if we take a walk on a $D$-regular expander graph with second eigenvalue $\lambda$ and $D = O(1)$, then the $t$-step walk will amplify the bias to $((1 - \frac{C}{2 \log\log(\abs{G})}) + 2 \lambda)^{\floor{t/2}}$. For this quantity to be at most $0.1$, it suffices to set $t > \frac{\log\log(\abs{G})}{C}(1 + 2\lambda) = \Theta(\log\log(\abs{G}))$. 

Therefore, after $t$ steps we obtain a generating set $S_0 \subset G^n$ with bias $0.1$, whose size is $\abs{S_0} \cdot D^t = O(\frac{n \log(\abs{G})^{2}}{(0.1)^{O(1)}} \cdot 2^{\Theta(\log\log(\abs{G}))}) = O(n (\log(\abs{G}))^{O(1)})$. 
\end{proof}

We remark that for the constant-error regime, this construction obtains almost the same parameters as that of \cite{arvind-permutation-groups}, who deal with the case of $\ZZ_d^n$. Their construction can be easily extended to any arbitrary abelian group via the projection lemma of \cite{arvind-rpp}, although it will have additional low-order terms in the size of the generating set. 

Next, the algorithm performs a wide replacement walk. We must specify the inner and outer graphs as well as the number of steps. Our parameters are almost identical to \cite{ta2017explicit}. We include them here for completeness.  

Let $\alpha = \Theta((\frac{\log\log(\frac{1}{\eps})}{\log(\frac{1}{\eps})})^{1/3})$. We will show that the wide replacement walk amplifies bias to $\eps$ and produces a generating set of size $O(\frac{n\log(\abs{G})^{O(1)}}{\eps^{2 + O(\alpha)}}) = O(\frac{n\log(\abs{G})^{O(1)}}{\eps^{2 + o(1)}})$. 

Let the ``width'' $s = \frac{1}{\alpha}$. 

\textbf{Inner Graph}: Let $D_2$ be the least power of two such that 
$D_2 \geq s^{4s}$. Let $b_2 = 4s \sqrt{2}\log(D_2)$. Let $D_1 = D_2^4$. Let $m = \log(D_1)$. 

Let $H = Cay(\ZZ_{2}^{ms}, A)$ for a generating set of size $\abs{A} = D_2$ (found, e.g via \cite{ta2017explicit}) 
such that the second eigenvalue is $\lambda(H) = \frac{b_2}{\sqrt{D_2}}$. 

\textbf{Outer graph}: Let $D_1 = D_2^4$. Find a $D_1$-regular expander graph $\Gamma$ with $\lambda(\Gamma) = \Theta(\frac{1}{\sqrt{D_1}})$ (using, e.g. \cite{alon-expanders-2020}). Identify its vertices with the $\eps_0$-biased set $S_{init}$. 

\textbf{Walk length}: Finally, set $t$ to be the least integer such that $\lambda(H)^{(1-4\alpha)(1-\alpha)t} \leq \eps$ and $t \geq \frac{s}{\alpha}$.  

\begin{prop}
The $t$-step wide replacement walk distribution is $\eps$-biased. 
\end{prop}

\begin{proof}
The bias after $t$ steps is given by $(\lambda(H)^s + s \lambda(H)^{s-1} + s^2 \lambda(H)^{s-2})^{\floor*{t/s}}$. Therefore, 

\begin{eqnarray}
(\lambda(H)^s + s \lambda(H)^{s-1} + s^2 \lambda(H)^{s-2})^{\floor*{t/s}} \leq (2s^2 \lambda(H)^{s-3})^{\floor*{t/s}} & \frac{\lambda(H) + s}{\lambda(H)^2} \leq s^2 \\
\leq (2s^2 \lambda(H)^{s-3})^{t/s - 1} \\
\leq (\lambda(H)^{s-4})^{t/s - 1} & 2s^2 \leq \lambda(H)^{-1} \\
= \lambda(H)^{\frac{s-4}{s} (t-s)} \\
= \lambda(H)^{(1 - \frac{4}{s}) (1 - \frac{s}{t}) t} \\
\leq \lambda(H)^{(1-4\alpha)(1-\alpha)t} & s = \frac{1}{\alpha}, t \geq \frac{s}{\alpha} \\ 
\leq \eps 
\end{eqnarray}

The last step follows by assumption on $t$.
\end{proof}

\begin{prop}
The support size of the wide replacement walk distribution is $O(\abs{S_{init}} \cdot \frac{1}{\eps^{2 + O(\alpha)}})$, where $S_{init}$ is the initial constant-bias set. 
\end{prop}

\begin{proof}
Recall that we identify our initial $0.1$-biased distribution with the vertices of the outer graph $\Gamma$. Therefore $N_1 = \abs{V(\Gamma)} = O(\frac{n \log(\abs{G})^{O(1)}}{\eps_0^{c}})$ for constant $\eps_0, c > 0$. Since $\eps_0$ is constant we can assume $D_2 \geq \eps_0^{-1}$.  
The walk begins at a uniform vertex of the replacement product, so the initial support size is $N_1 N_2$. After $t$ steps it increases by a factor of $D_2^t$. Therefore 

\begin{eqnarray}
N_1 N_2 D_2^t = O(\frac{n \log(\abs{G})^{O(1)}}{\eps_0^c}  N_2 D_2^t) \\
= O(\frac{n \log(\abs{G})^{O(1)}}{\eps_0^c} D_2^{4s} D_2^t) \\
= O(n \log(\abs{G})^{O(1)} \cdot D_2^{4s + t + c}) & \eps_0^{-1} \leq D_2 \\
\leq O(n \log(\abs{G})^{O(1)}\cdot D_2^{4 \alpha t + t + c}) & s \leq \alpha t \\
\leq O(n \log(\abs{G})^{O(1)} \cdot D_2^{t (1 + 5\alpha)}) & c \leq \alpha t \\
\end{eqnarray}

Next, notice $b_2 = 4 \sqrt{2} s \log(D_2) = 4 \sqrt{2} \cdot 4s^2 \log(s) \leq s^4$ for sufficiently large $s$ (equivalently, small enough $\eps$). Therefore, $D_2 \geq (s^4)^s \geq b_2^s = b_2^{1/\alpha}$. Therefore $D_2^{1/2 - \alpha} \leq \lambda(H)^{-1} = \frac{\sqrt{D_2}}{b_2}$. 

It follows that 

\begin{eqnarray}
D_2^{t} \leq (\lambda(H)^{-1})^{\frac{t}{1/2 - \alpha}} \\
= (\lambda(H)^{-1})^{\frac{2t}{1 - 2 \alpha}} \\
= (\eps^{-1})^{\frac{1}{(1-4\alpha)(1-\alpha)t} \frac{2t}{1-2\alpha}} \\
\leq (\eps^{-1})^{2(1 + 8\alpha)}
\end{eqnarray}

The last inequality follows for small enough $\alpha$ (equivalently, small enough $\eps$). 

Finally, 

\begin{eqnarray}
D_2^{t(1 + 5\alpha)} \leq (\eps^{-1})^{2(1 + 8\alpha)(1 + 5\alpha)} \\
\leq (\eps^{-1})^{2(1 + 14\alpha)}
\end{eqnarray}

Therefore, our overall support size is $O(\frac{n\log(\abs{G})^{O(1)}}{\eps^{2 + O(\alpha)}})$. In particular, since $\alpha \to 0$ as $\eps \to 0$, the support size is $O(\frac{n \log(\abs{G})^{O(1)}}{\eps^{2 + o(1)}})$. 
\end{proof}


%% file: applications.tex

\section{Applications}

In this section, we will demonstrate the algorithmic applications of our construction of expanding generating sets for abelian Cayley graphs. As before let $G$ denote a finite abelian group and $n \geq 1$ an integer. 

\subsection{Almost k-wise independence}

Let $D \sim G^n$ and $U$ denote the uniform distribution on $G^n$. 
We say that $D$ is $(\eps, k)$-wise independent 
if for every $I \subset [n]$ of size $k$, the restriction of $D$ onto $I$-indices, denoted $D_I$, is $\eps$-close to $U_I$ in statistical distance. 

Let $\Delta$ denote statistical distance. Vazirani's XOR Lemma asserts that if $G = \FF_2$ and $D$ is $\eps$-biased, then $\Delta(D, U) \leq \eps \cdot \sqrt{2^n}$. In other words, if $D$ is near-uniform in a weak sense (namely, if $D$ is $\eps$-biased), then $D$ is also near-uniform in a strong sense (with respect to statistical distance), at the cost of a $\sqrt{2^n}$ factor. 

From the proof of the lemma, it is easy to see that $D$ is also $(\eps \cdot \sqrt{2^k}, k)$-wise independent for every $k \leq n$. 

Vazirani's XOR Lemma generalizes straightforwardly to the case of an arbitrary abelian group $G$. For the sake of completeness we include the proof here. 

First, we need a lemma concerning Fourier coefficients of the uniform distribution. 

\begin{lemma}
Let $D \sim G^n$ be an arbitrary distrubtion and $U \sim G^n$ be the uniform distribution. 

(i) Let $\chi: G^n \rar \CC^*$ be the trivial character. Then $\EE_{x \sim D}[\chi(x)] = 1$. 

(ii) For any nontrivial character $\chi: G^n \rar \CC$, $\EE_{x \sim U}[\chi(x)] = 0$. 
\end{lemma}

\begin{proof}
(i) By definition $\chi(g) = 1$ for any $g \in G^n$, so $\EE_{x \sim D}[\chi(x)] = 1$ for any $D$. 

(ii) First, consider the special case $G = \ZZ_d$ for some $d \geq 2$. Then a nontrival $\chi: G^n \rar \CC$ corresponds to some $a \in \ZZ_d^n \setminus \{\vec{0}\}$. Observe that $\la x, U \ra$ is uniform on $\ZZ_d$, where the inner product is taken modulo $d$. Therefore $\EE_{x \sim U}[\chi(x)] = \EE_{x \sim U}[exp(\frac{2\pi i}{d} \la a, x \ra)] = 0$. 

Next, for arbitrary abelian $G$, observe that $G = \ZZ_{d_1} \oplus \cdots \oplus \ZZ_{d_k}$. Any nontrivial character $\chi$ on $G^n$ is a product of characters on the factor groups $\ZZ_{d_i}$. We have already shown each of these characters has expectation $0$ on uniform inputs. 
The result follows from $k = 1$ case. 
\end{proof}

Next, we can prove our claim. 

\begin{prop}[Generalized Vazirani XOR Lemma]
Let $D \sim G^n$ be $\eps$-biased and $U$ denote the uniform distribution on $G^n$. Then 
$$\Delta(D, U) \leq \eps \cdot \sqrt{\abs{G}^n}$$
\end{prop} 

\begin{proof}
Let $T: G^n \rar \CC$ be the (normalized) indicator function of some arbitrary statistical test (that is, an event on the outcome space $G^n$). Then, writing the distance in the Fourier basis, 

\begin{align}
\Delta(D, U) &= \abs{\EE_D(T) - \EE_U(T)} \\
&= \abs{\sum\limits_{\chi \in \hat{G^n}} \hat{T}_{\chi} (\EE_D[\chi] - \EE_U[\chi])} \\
&\leq \sum\limits_{\chi \in \hat{G^n}} \abs{\hat{T}_{\chi}} \abs{\EE_D[\chi] - \EE_U[\chi])} \\
&\leq \sum\limits_{\chi \text{ trivial}}  \abs{\hat{T}_{\chi}} \cdot \abs{\EE_D[\chi] - \EE_U[\chi])} + \sum\limits_{\chi \text{ nontrivial}}  \abs{\hat{T}_{\chi}} \cdot \abs{\EE_D[\chi] - \EE_U[\chi])} \\
&\leq \abs{\hat{T}_{\chi}} \cdot \abs{0 - 0} + \sum\limits_{\chi \text{ nontrivial}}  \abs{\hat{T}_{\chi} \cdot  \abs{\eps - 0}} \\
&\leq \eps \sum\limits_{\chi \text{ nontrivial}}  \abs{\hat{T}_{\chi}} \\
&\leq \eps \sqrt{\abs{G}^n} \sqrt{\sum\limits_{\chi \text{ nontrivial}}  (\hat{T}_{\chi})^2} \\
&\leq \eps \sqrt{\abs{G}^n}
\end{align}

The last step is due to Plancherel's Theorem. 
\end{proof}

A special case of this fact is shown in \cite{amn} Theorem 4.5, when $G$ is replaced with a finite field of prime order. 

As a corollary, an $\eps$-biased distribution has statistical distance at most $\eps \cdot \sqrt{\abs{G}^k}$ from a $k$-wise independent distribution, since the latter is uniform on sets $k$ indices, and a restriction of an $\eps$-biased set to any subset of indices is still $\eps$-biased. Therefore, to obtain an $(\eps, k)$-wise independent distribution we simply construct a distribution with bias $\frac{\eps}{\sqrt{\abs{G}^k}}$. This requires a support size of $O(\frac{n \log{\abs{G}}^{O(1)} \cdot \abs{G}^{k + o(1)}}{\eps^{2 + o(1)}}) = O(\frac{n \cdot \abs{G}^{k + o(1)}}{\eps^{2 + o(1)}})$. The following proposition follows immediately. 

\begin{prop}[Almost $k$-wise independent sets over abelian groups]
Let $G$ be a finite abelian group, and $k \leq n$ be positive integers. For any input $\eps > 0$ and generating set of $G$, there exists a deterministic, polynomial-time algorithm whose output is an $(\eps, k)$-wise independent distribution over $G^n$. The support size is $O(\frac{n \cdot \abs{G}^{k + o(1)}}{\eps^{2 + o(1)}})$
\end{prop} 

\subsection{Remote Point Problem}

Let $G$ be a group and $H \leq G^n$ a subgroup given by some generating set $H^\prime \subset H$. For a given $G, H$ and integer $r > 0$, the Remote Point Problem is to find a point $x \in G^n$ such that $x$ has Hamming distance $> r$ from all $h \in H$, or else reject. 


Alon, Panigrahy, and Yekhamin introduced the Remote Point Problem over $G = \FF_2$ \cite{alon-2009-rpp}. Later, Arvind and Srinivasan generalized the problem to any group, and extended the algorithm of Alon, Panigrahy, and Yekhamin to the generalized setting \cite{arvind-rpp}. 

Their algorithm proceeds in essentially three steps. For a subgroup $H \leq G^n$, let its dimension be $dim(H) \defeq \log_{\abs{G}}(H)$.  

Upon an input $H \leq G^n$ of dimesion $k \leq n/2$, 

(1) Compute subgroups $H_1, ..., H_m$ which cover $H$, each of which has dimension $\leq 2n/3$. We obtain $m = n^{O(c)}$ for a constant $c$ which controls how good our output distance is. 

(2) Construct a symmetric multiset $S \subset G^n$ such that the Cayley graph $Cay(G^n, S)$ has $\lambda(Cay(G^n, S)) \leq \alpha$. They require $\alpha \leq \frac{1}{m^2} \approx \frac{1}{n^{20 c}}$. 

(3) Exhaustively search $S$ for a point which is outside of $\bigcup_i H_i$. Return that point $s$. 

Let $d_H$ denote Hamming distance on $G^n$. From arguments in \cite{arvind-rpp} it follows that $d_{H}(x, H) \geq \frac{c n \log(k)}{k}$ for dimension $k$. 

The size of $\abs{S}$ is a bottleneck in the algorithm of \cite{arvind-rpp}. In general they obtain expanding generating sets of size $O((\log(\abs{G}) + \frac{n^2}{\eps^2})^5)$, and for $\log(\abs{G}) \leq \log(\frac{n^2}{\eps^2})^{O(1)}$ this is improved to $O(\frac{n^2}{\eps^2})$. Therefore, we shrink the support size of $S$ by a factor of at least $\Tilde{O}(n)$ and hence speed up the exhaustive search step (3). 

\subsection{Randomness-Efficient Low-Degree Testing}

Let $\FF_q$ be a finite field of $q$ elements, $n, d \geq 1$, and $f: \FF_q^n \rar \FF_q$. The low-degree testing problem is to determine whether $f$ is a degree-$d$ polynomial or far from all such polynomials in Hamming distnace. 

In the line oracle model, a tester is given query access to the function $f$, along with a line function $g$. Let $\lines$ denote all lines $\{\vec{a} + t\vec{b}: t \in \FF\} \subset \FF^n$, where $\vec{a}, \vec{b} \in \FF^n$. In general there are multiple distinct choices of $\vec{a}, \vec{b}$ which may describe the same line, so we implicitly fix some parametrization of $\lines$. 

Given a description of a line, the line oracle $g$ returns a univariate polynomial of degree $d$ defined on that line. Hence we write $g: \lines \rar \FF[t]$, where the image of $g$ is understood to only contain degree-$d$ polynomials. 

The Rubinfeld-Sudan test now proceeds as follows \cite{rubinfeld-sudan-1996}. 

If $f$ is indeed a degree-$d$ polynomial, then one can set $g(\ell) = f\vert_{\ell}$ for all $\ell \in \lines$, and the following two-query test clearly accepts. 

(i) Select $x, y \in \FF^n$ independently, uniformly at random.

(ii) Let $\ell$ be the line determined by $\{x + ty: t \in \FF_q\}$. Accept iff $f(x)$ agrees with $g(\ell)(x)$. 

Ben-Sasson et al derandomized the line-point test as follows. Their algorithm flips a fair coin. If heads, it samples $y$ from an $\eps$-biased set $S \subset \FF_q^n$ rather than from the entire space $\FF_q^n$. If tails, it checks whether $f(x)$ agrees with $g(\ell)(x)$, where $\ell = \{0 + tx: t \in \FF_q\}$. We call this the ``derandomized line-point test.''

Our construction improves the randomness-efficiency of the test since the $\eps$-biased space $S$ is smaller. The soundness parameters of the test are the same. 

\begin{prop}[Improved \cite{vadhan-pcp} Theorem 4.1]
Let $\FF_q$ be a finite field of $q$ elements, $n \geq 1$,  $f: \FF_q^n \rar \FF_q$, $g: \lines \rar \FF_q[t]$. The derandomized line-point point test has sample space size $O(q^n \cdot \frac{n\log(q)^{O(1)}}{\eps^{2 + o(1)}})$. Further, there exists a universal constant $\alpha > 0$ such that for $d \leq q/3$, $~n \leq \frac{\alpha q}{\log(q)}, \eps < \frac{\alpha}{n \log(q)}, \delta \leq \alpha$, if the  derandomized line-point test accepts with probability $\geq 1- \delta$ then $f$ has Hamming distance at most $4\delta$ from a degree $d$ polynomial. 
\end{prop}

\subsection{Randomness-Efficient Verification of Matrix Multiplication}

Suppose $A, B, C$ are $(n \times n)$ matrices whose entries belong to some finite field $\FF_q$ or cyclic group $\ZZ_q$ for $q \geq 2$. Let $R$ denote either $\FF_q$ or $\ZZ_q$. We wish to verify whether $AB = C$ over $R$. 

Naively we can multiply $A, B$ and check entry-wise, but this takes $O(n^{\omega})$ time, where $\omega \approx 2.37$ is the exponent of matrix multiplication \cite{laser-2021}. 

A simple randomized algorithm is to sample vectors $x \in R^n$ uniformly, and then check whether $ABx = Cx$ \cite{freivalds1977}. This requires three matrix-vector multiplications, which takes $O(n^2)$ time \footnote{For simplicity we consider addition and multiplication over $R$ to be constant-time operations. Our results do not depend on the details of implementing arithmetic over $R$.}. 

Our algorithm replaces these uniform samples with samples from a small-bias set. 
Observe that if $AB \neq C$, then the probability that $\PP_{x \sim R^n}[ABx = Cx] = \PP_{x \sim R^n}[(AB - C)x = \vec{0}] = \frac{1}{q}$. If $S \subset R^n$ is $\alpha$-biased then $\PP_{x \sim S}[ABx = Cx | AB \neq C] \leq \frac 1 q + \alpha$. 

\begin{prop}
Let $R$ denote a finite field $\FF_q$ or cyclic group $\ZZ / q\ZZ$. Given matrices $A, B, C \in R^{n \times n}$ and $\alpha$-biased set $S \subset R^n$, there exists an $O(n^2)$ time randomized algorithm to decide whether $AB = C$ with one-sided error $\frac{1}{q} + \alpha$. It uses $O(\log(\frac{n\log(q)^{O(1)}}{\alpha^{2 + o(1)}}))$ random bits. 
\end{prop}

